\documentclass[aip,graphicx]{revtex4-1}
\usepackage{mathrsfs}
\usepackage{graphicx}
\usepackage{amsmath}
\usepackage{epstopdf}
\usepackage{appendix}
\usepackage{ulem}
\usepackage{color}
\usepackage{bm}
\usepackage{gensymb}

\usepackage{amsfonts}

\usepackage{comment}

\newcommand{\beq}{\begin{equation}}
\newcommand{\eeq}{\end{equation}}
\newcommand{\bea}{\begin{eqnarray}}
\newcommand{\eea}{\end{eqnarray}}

\def\ang{\,\textrm{\AA}}
\def\bath{\textrm{bath}}

\def\T{\,\textrm{T}}

\def\MHz{\,\textrm{MHz}}
\def\GHz{\,\textrm{GHz}}
\def\mus{\,\mu\textrm{s}}

\def\Tr{\mathop{\textrm{Tr}}}

\usepackage{graphicx}% Include figure files
\usepackage{dcolumn}% Align table columns on decimal point
\usepackage{bm}% bold math
\usepackage[utf8]{inputenc}
\usepackage[T1]{fontenc}
\usepackage{mathptmx}
\usepackage{etoolbox}

\draft % marks overfull lines with a black rule on the right

\begin{document}

%\preprint{}

\title{Simulating decoherence of two {\color{black} coupled} spins using the generalized cluster correlation expansion} %Title of paper



%\author{}
%\thanks{}
%\affiliation{}

\author{Xiao Chen}
\affiliation{Department of Physics, Northeastern University, Boston MA 02115 USA}

\author{Silas Hoffman}
\affiliation{Department of Physics,  University of Florida, Gainesville FL 32611 USA}

\author{James N. Fry}
\affiliation{Department of Physics, Northeastern University, Boston MA 02115 USA}

\author{Hai-Ping Cheng}
\email{ha.cheng@northeastern.edu}
\affiliation{Department of Physics, Northeastern University, Boston MA 02115 USA}


\date{\today}

\begin{abstract}
We simulate the coherence of two coupled electron spins interacting with a bath of nuclei using the generalized cluster correlation expansion (gCCE) method. 
An exchange interaction between the electrons facilitates a family of entangling gates that can be spoiled by nuclear-induced dephasing.
Consequently, we study the dephasing of the coherent two-electron system by characterizing the $T_2$ and $T_2^*$ of the two-electron reduced density matrix for various system parameters {\color{black} in the range} mimicking magnetic molecules, {\color{black} including magnetic field strength and orientation, exchange interaction strength, distance between the two spins, minimum distance between electron and nuclei and between nuclei, and nuclei density}. {\color{black} We find the optimal regime for each parameter in which the coherence time is maximized and provide a physical understanding of it.}
\end{abstract}


\maketitle %\maketitle must follow title, authors, abstract and \pacs

%\section{}
%\label{}
%\subsection{}
%\subsubsection{}

% If in two-column mode, this environment will change to single-column format so that long equations can be displayed. 
%\begin{widetext}
%$$\mbox{put long equation here}$$
%\end{widetext}

% Figures should be put into the text as floats. 
% Use the graphics or graphicx packages (distributed with LaTeX2e).
% Put the label that you will use with \ref{} command in the braces of the \label{} command.
%
% \begin{figure}
% \end{figure}

% \begin{table}
% \end{tabular}
% \end{table}

\section{Introduction}

Localized electrons have been well studied theoretically and experimentally as candidates to store and manipulate quantum information.\cite{burkard2023semiconductor,vandersypen2017interfacing} In addition to being natural qubit candidates, owing to their two-level structure, they are also endowed with the ability to interact. Specifically, the exchange interaction facilitates strong, short-range interactions while their magnetic dipoles allow for weaker, long-range interactions. Importantly, when using electrons as qubits, these interactions facilitate entangling two-qubit gates, which are necessary for universal quantum computation.\cite{petta2005coherent,nowack2011single,veldhorst2015two,watson2018programmable,zajac2018resonantly} Moreover, in multielectron encodings of qubits, these entangling gates are often necessary to enable single-qubit control.\cite{levy2002universal,petta2005coherent,bacon2000universal,divincenzo2000universal,kempe2001theory} Consequently, understanding the decoherence of two interacting electron spins is an important problem for achieving high-fidelity two-spin-qubit gates and single-qubit gates on multielectron encoded qubits.

One of the dominant sources of decoherence at low temperatures in spin-qubit systems is due to the surrounding nuclear spin bath.\cite{yao2006theory,khaetskii2002electron}
The state-of-the-art method to simulate decoherence of coupled  {\color{black} electron} spins due to a nuclear spin bath is the generalized cluster-correlation expansion (gCCE)\cite{yang2020longitudinal,onizhuk2021probing,yang2016quantum,RevModPhys.97.021001}, 
{\color{black} in which full dynamics of the central-spin system is incorporated to better account for population change and virtual electron spin flip-flop processes that are crucial for decoherence at small fields and at clock transitions.}
The majority of the  {\color{black} gCCE} literature  {\color{black} simulating multi-central-spin decoherence focus} on decoherence of coupled electron-nuclear spin qubits\cite{balian2014quantum,onizhuk2023bath,maile2024performance}.
{\color{black} Works have been done to characterize decay of entangled states of two electron spins that are uncoupled\cite{bragar2015dynamics} or weakly coupled by undesired dipolar interactions\cite{kwiatkowski2018decoherence}, with other methods such as conventional cluster-correlation expansion.}  {\color{black} gCCE} simulation of decoherence of {\color{black} strongly} coupled two-electron spin systems due to quantum noises from nuclear spins, on the other hand, is still a largely unexplored subject.

In this work, we study precisely a model of two coupled spin-$\frac{1}{2}$ electrons embedded in a random nuclear spin bath. 
{\color{black}
We focus on the case where the interaction between electron spins is an isotropic exchange interaction.
}
In particular, using gCCE, we calculate the evolution of the two-spin reduced density matrix (RDM) {\color{black} in pure dephasing regime} under free induction and upon application of  {\color{black} the Hahn-echo} pulse sequence. 
{\color{black} The decay of off-diagonal elements of the RDM, which reflects decoherence in the two-spin system, is closely related to many experimentally measured quantities including gate fidelity, if the coupled two spins are utilized as a two-qubit entangling gate, to the decay of transverse magnetization in spin echo measured in a standard Hahn-echo experiment in pulsed Electron Paramagnetic Resonance (EPR) and to the decoherence of two-spin encoded single qubits, etc.}
Under a large range of system parameters, but guided by those typical for magnetic molecular spin systems in pure dephasing regime, we focus on {\color{black} time evolution} of the off-diagonal elements. 
We find an optimal regime of {\color{black} each} parameter in which the coherence time is maximized{\color{black}, while other parameters are fixed}. 
{\color{black} In our model of random nuclear bath, aligning the external magnetic field to the line joining the two electron spins generally leads to longer coherence times than the orthogonal orientation. 
The coherence times increase with the strength of the external field until the regime of large fields, where the coherence times remain constant. 
Changing the strength of the exchange interaction has essentially no effect on the decoherence.
An optimal distance between the electron spins for maximum coherence times is found to be comparable to the minimum electron-nuclei distance.
Increasing the minimum electron-nuclei distance or decreasing the density of nuclear spins improves coherence.
Increasing the minimum spacing between nuclear spins has little effect on free induction decay but prolongs the Hahn-echo coherence times.
}

The rest of the paper is organized as follows. In Sect.~\ref{sec:model and gCCE}, we introduce our model,  consisting of  {\color{black} two coupled electron spins} interacting with a random nuclear spin bath, %A brief description of the gCCE method is also included in this section. 
{\color{black} and apply the so-called pair-correlation approximation to the two-spin problem in large field and exchange interaction which guides our physical understanding of the gCCE results.}
In Sec.~\ref{sec:results_and_discussion} we present the calculated coherence {\color{black} time of the off-diagonal elements of the two-electron RDM,}   $T_2${\color{black}, for the case of Hahn-echo pulse sequence,} and $T_2^*${\color{black}, for the case of free induction decay.} 
The effect of  {\color{black} the aforementioned} physical parameters on decoherence {\color{black} and a physical understanding of the dependence of decoherence times on these parameters are provided in detail.} The implications on the experimental design of {\color{black} realistic} magnetic molecular two-spin systems with longer coherence time are also discussed.
{\color{black}
Finally, we discuss the change in the dephasing dynamics and show the failure of the gCCE method when a non-negligible magnetic dipolar interaction between the two electron spins is included. 
}
In Sect.~\ref{sec:Conclusion} we conclude the paper.

\section{Model and theoretical analysis}\label{sec:model and gCCE}

\subsection{Model of two-spin decoherence}\label{subsec:model}
\begin{figure}[htp]
$\begin{array}{c}
\includegraphics[width=0.6\columnwidth]{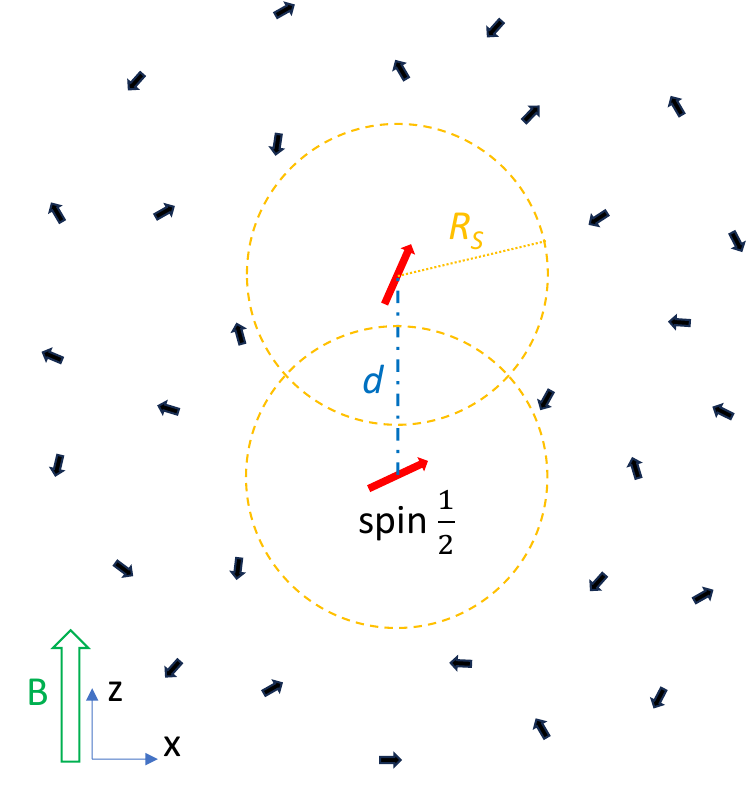}
\end{array}$\caption{A sketch of the model. The electron spins and the nuclear bath spins are represented by the red and black arrows, respectively.
}\label{fig:model_sketch}
\end{figure}

{\color{black} The exchange interactions between magnetic centers in molecules usually take the form of double or superexchange mediated by linker atoms between the centers which are typically much stronger than the magnetic dipolar interaction even when the spin centers are many covalent bonds apart\cite{mironov1998generalized,JESCHKE2015}. We focus on the situation where the exchange interaction is dominant and assume that it is the only interaction between electron spins.
The effect of dipolar interaction will be discussed in Sect.~\ref{subsec:dipolar}.} 
In order to study decoherence dominated by nuclei induced magnetic fluctuations in molecular magnetic systems, we consider two electron spin-$1/2$'s coupled by an isotropic exchange interaction, embedded in a bath of  nuclear spins and in an external magnetic field (Fig.~\ref{fig:model_sketch}) which is described by the Hamiltonian
\beq
\hat{H}=\hat{H}_{S}+\hat{H}_{B}+\hat{H}_{SB}\,.\label{eq:total Hamiltonian}
\eeq
The electron Hamiltonian {\color{black}in unit of frequency}  is
\bea
\hat{H}_{S}&=&-\gamma_{e}B\hat{S}^z_{1}-\gamma_{e}B\hat{S}^z_{2}+J\hat{\vec{S}}_{1}\cdot\hat{\vec{S}}_{2}\,.\label{eq:H_S}
\eea
This describes two electrons coupled by an isotropic exchange interaction of magnitude $J$ in a magnetic field of magnitude $B$ which we have chosen to be oriented along the $z$ axis without loss of generality. Here, $\hat{\vec{S}}_j$ is the vector of dimensionless spin operators associated with the $j$th spin-1/2 electron and $\gamma_e$ is the bare electron gyromagnetic ratio.  The nuclear bath Hamiltonian is
\bea
\hat{H}_{B}&=&-\gamma_{p}B\sum_{n}\hat{I}^z_{n}+\sum_{n<m}\hat{\vec{I}}_{n}\cdot\overleftrightarrow{D}_{nm}\cdot\hat{\vec{I}}_{m}\,.\label{eq:H_B}
\eea
While our formalism could be generalized to dephasing due to any nuclei, we consider hydrogen nuclei, i.e. protons, because they {\color{black} are usually the dominant decoherence source}  in magnetic molecular systems.\cite{chen2024spin,zecevic1998dephasing,canarie2020quantitative,chen2020decoherence} Hence, $\gamma_p$ is the proton gyromagnetic ratio and $\hat{\vec{I}}_n$ is the vector of dimensionless spin-1/2 operators for the $n$th proton. The magnetic point dipolar interactions between the nuclear spins are
\bea
\overleftrightarrow{D}_{nm}&=&-\gamma_{p}^2\frac{\mu_{0}\hbar}{2|\vec{r}_{nm}|^{5}}\left[3\vec{r}_{nm}\otimes\vec{r}_{nm}-|\vec{r}_{nm}|^{2}\mathbb{I}_{3\times3}\right] ,\,\label{eq:p-p dipolar}
\eea
where $\vec{r}_{nm}$ is the real space vector joining the positions of the $n$th and $m$th nuclei, $\mathbb{I}_{3\times3}$ is the three-dimensional identity matrix, and $\mu_0$ is the vaccuum permeability. The electrons and nuclei interact with each other via the hyperfine interaction according to
\bea
\hat{H}_{SB}&=&\sum_{n}\hat{\vec{S}}_{1}\cdot\overleftrightarrow{A}_{1n}\cdot\hat{\vec{I}}_{n}+\sum_{n}\hat{\vec{S}}_{2}\cdot\overleftrightarrow{A}_{2n}\cdot\hat{\vec{I}}_{n},\label{H_SB}
\eea
where the hyperfine tensor between the $j$th electron and $n$th spin is assumed dipolar
\bea
\overleftrightarrow{A}_{jn}&=&-\gamma_{e}\gamma_{p}\frac{\mu_{0}\hbar}{2|\vec{r}_{jn}|^{5}}\left[3\vec{r}_{jn}\otimes\vec{r}_{jn}-|\vec{r}_{jn}|^{2}\mathbb{I}_{3\times3}\right] , \,
\label{eq:dipolar_hyperfine}
\eea
where $\vec{r}_{jn}$ is the real space vector joining the positions of the $j$th electron and the $n$th nuclei.
{\color{black} We focus on magnetic molecules in which protons do not overlap significantly with the spin density or the exchange pathway, which include transition-metal complexes and lanthanide-based complexes. Therefore, the Fermi contact part of the hyperfine interaction is ignored.}

The electrons are positioned a distance $d$ away from each other either along the direction of the magnetic field, i.e. $\vec d=d \hat z$,  or perpendicular to the magnetic field direction, i.e. $\vec d=d\hat x$. In molecular magnetic systems, repulsive electrostatic forces typically force both protons away from each other, and the hydrogen atoms are usually not directly bonded to the atoms hosting the electron spins. Consequently, we enforce a minimum distance of $|\vec r_{jn}|=R_S$ and $|\vec r_{nm}|=R_B$ for the electron-proton and proton-proton distances, respectively. Aside from this restriction, the protons are randomly positioned in real space with equal probability to a density of $n_B$. 

It is convenient to recall that the eigenstates of the isolated electron system [Eq.~(\ref{eq:H_S})] are a spin singlet and three spin triplets,
\bea
\left|E_{S}\right\rangle &=&\frac{1}{\sqrt{2}} \, (\left|\uparrow_1\rangle|\downarrow_2\right\rangle-\left|\downarrow_1\rangle|\uparrow_2\right\rangle),\nonumber\\
\left|E_{-1}\right\rangle &=&\left|\downarrow_1\rangle|\downarrow_2\right\rangle,\nonumber\\
\left|E_{0}\right\rangle &=&\frac{1}{\sqrt{2}} \, (\left|\uparrow_1\rangle|\downarrow_2\right\rangle+\left|\downarrow_1\rangle|\uparrow_2\right\rangle),\nonumber\\
\left|E_{1}\right\rangle&=&\left|\uparrow_1\rangle|\uparrow_2\right\rangle\,,\label{eq:isotropic_exchange_eigenenergy_basis}
\eea
respectively, and $|\uparrow_j\rangle$ ($|\downarrow_j\rangle$) corresponds to the positive (negative) eigenstate of $\hat S^z_j$. The respective eigenvalues are 
\bea
E_{S}&=&-\textstyle\frac{3}{4}J\,, \nonumber\\
E_{-1}&=& B_{z}\gamma_{e}+\textstyle\frac{1}{4}J\,, \nonumber\\
E_{0}&=&\textstyle\frac{1}{4}J\,, \nonumber\\
E_{1}&=& -B_{z}\gamma_{e}+\textstyle\frac{1}{4}J\,. 
\label{eq:isotropic_exchange_two-qubit_energies}
\eea
In typical molecular magnetic systems, $0.1\GHz\leq J \leq 10^3\GHz$, $3\ang\leq R_S \leq 10\ang$, $0.001/\ang^3\leq n_B \leq 0.02/\ang^3$, $1\ang\leq R_B \leq 5\ang$, $0.01\T\leq |\vec{B}| \leq 10\T$ which corresponds to a large difference in level spacing of the two-electron system [Eq.~(\ref{eq:isotropic_exchange_two-qubit_energies})] and the nuclear Zeeman energy compared to the electon-nuclear dipole energy, i.e. $\gamma_{e}\gamma_{p}\mu_{0}\hbar/2|\vec{r}_{jn}|^{3}\ll|\gamma_p B-|E_\alpha-E_\beta||$ for $\alpha,\beta=S,-1,0,1$ and $\alpha\neq\beta$. We {\color{black}only consider these parameter ranges in the model and} are consequently in the pure dephasing regime {\color{black} where the decoherence of the two interacting electron spins manifests itself as the decay of the off-diagonal elements of the reduced density matrix (RDM) of the electron spins in the basis of 
Eq.~(\ref{eq:isotropic_exchange_eigenenergy_basis}).}

\subsection{Decoherence}\label{subsec:Decoherence}

In general, we are interested in the off-diagonal elements of the reduced density matrix after the system has undergone time evolution with $N$ dynamical decoupling pulses applied. We focus on the scenarios $N=0$, which corresponds to free induction decay (FID), and $N=1$, which corresponds to Hahn echo experiments. The full density matrix {\color{black} at the end time $t$ of the pulse sequence} is $\rho_N(t)=\hat U_N\hat\rho(0)\hat U_N^\dagger$
with
\bea
\hat{U}_N&=&\exp(-i\hat{H}\tau)\left[\pi_{xx}\exp(-i\hat{H}\tau)\right]^N\,.\label{eq:unitary_evolution}
\eea
Here, $\exp(-i\hat{H}\tau)$ is the free evolution of the full electron-proton system [Eq.~(\ref{eq:total Hamiltonian})], $\pi_{xx}=\exp[-i\pi(\hat{S}_{1x}+\hat{S}_{2x})]$ is an operational expression for a simultaneous $\pi$-pulse on each electron, and $\tau=t/(N+1)$. {\color{black} Note that since Hamiltonians have been chosen to have the unit of linear frequency, an implicit factor of $2\pi$ exist in the power of propagators like $\exp(-i\hat{H}\tau)$, which is not shown for clarity.} 
Assuming that at $t=0$ the electron and bath system are uncorrelated, $\hat{\rho}(0)=\hat{\rho}_{S}(0)\otimes\hat{\rho}_{B}(0)$, coherence in this system is conveniently quantified by the normalized off-diagonal elements of the reduced density matrix,
\beq
L^N_{\alpha\beta}(t)=\left|\frac{\langle E_\alpha|\textrm{Tr}_{B}[\hat{\rho}_N(t)]|E_\beta\rangle}{\langle E_\alpha|\hat{\rho}_{S}(0)|E_\beta\rangle}\right|\,,\label{eq:coherence_function}
\eeq
where $\textrm{Tr}_{B}[\cdots]$ indicates the trace over the proton degrees of freedom.

Within the pure dephasing Hamiltonian of the form Eq.~(\ref{eq:pure_dephasing_total_H}) we find $L^0_{\alpha\beta}(t)=|\textrm{Tr}_{B}[e^{-i\hat{H}_{B;\alpha}t}\hat{\rho}_{B}(0)e^{i\hat{H}_{B;\beta}t}]|$ and $H_{B;\alpha}$ given by Eq.~(\ref{eq:QDPT_H}) above. Notice that this is independent of the initial state of the electron system conditioned on $\langle E_\alpha|\hat{\rho}_{S}(0)|E_\beta\rangle\neq0$; we find that $L^1_{\alpha\beta}(t)$ is similarly independent of the initial electronic spin configuration. Because the temperatures in typical magnetic molecular experiments are much higher than the Zeeman splitting of the protons, we take $\rho_B(0)=\mathbb{I}_{2^{N_B}\times 2^{N_B}}$, i.e. the completely mixed state of the protons{\color{black}, $N_B$ being the number of protons}.

gCCE as implemented in the PyCCE package\cite{onizhuk2021pycce} was used to numerically calculate the time evolution of the RDM of the two electron spins {\color{black} according to Eq.~\ref{eq:unitary_evolution} with the full spin Hamiltonian as in Eqs.~\ref{eq:total Hamiltonian}--\ref{eq:dipolar_hyperfine}}. 
%{Once the gCCE is convergent, it provides an accurate account of the population change of the central spin system and the central spin virtual flip-flop processes, the later of which is crucial for the description of decoherence at low fields even the central spin system evolution remains pure dephasing (no population change).}
See Supplementary Material Sect.~S1
for a brief description of the gCCE method
and Sect.~S2 for more computational details. 
In order to study the effect of a uniformly distributed nuclear bath with no bias towards any specific random spatial configuration generated as described in Sect.~\ref{subsec:model}, 
{\color{black}  unless stated otherwise,} we numerically calculate $L^N_{\alpha\beta}(t)$ over three hundred random spatial configurations of the proton bath and average the results together to arrive at the final results. {\color{black} The aim is to discover and understand parameters that affect coherence times of the two coupled electron spins in this random bath model, from which useful implications on increasing coherence time by tuning these parameters can be drawn for realistic magnetic molecular systems.}

{\color{black}
\subsection{Projected total Hamiltonians}
}
{\color{black} In order to gain a physical understanding of the parameter dependence of coherence times in full gCCE results, a first step is to consider an approximated total Hamiltonian [Eqs.~(\ref{eq:total Hamiltonian}) thru (\ref{eq:dipolar_hyperfine})] by projecting onto the electron energy states. The projected Hamiltonian consists of terms of separated electron-nuclei form, with the nuclear part clearly showing the interactions that govern the bath dynamics. It is well known that bath dynamics determines the electron spin decoherence.\cite{khaetskii2002electron,de2003theory}}
Denoting the identity operator in the state space of the electron (nuclear) spins by $\mathbb{I}_S$ ($\mathbb{I}_B$), the sum of the electon-nuclear interaction {\color{black} $\hat{H}_{SB}$} and the nuclear spin Hamiltonian $\mathbb{I}_{S}\otimes\hat{H}_{B}$ can be treated as a perturbation to the electron spin Hamiltonian $\hat{H}_S\otimes\mathbb{I}_B$ to obtain a projected total Hamiltonian {\color{black} within quasidegenerate perturbation theory (QDPT)\cite{lowdin1951note,luttinger1955motion}} of the form
\beq
\hat{H}{\color{black}_{ proj}}=\sum_{\alpha=S,-1,0,1}\left|E_{\alpha}\right\rangle \left\langle E_{\alpha}\right|\otimes\hat{H}_{B;{\alpha}},\label{eq:pure_dephasing_total_H}
\eeq
in which the effective nuclear spin Hamiltonian when the two electron spins are in state $|E_{\alpha}\rangle$ is
\bea
&&\hat{H}_{B;{\alpha}}=E_{\alpha}+\left\langle E_{\alpha}\right|\hat{H}_{SB}\left|E_{\alpha}\right\rangle +\hat{H}_{B}\nonumber\\
&&\quad +\sum_{\beta\neq \alpha}\frac{\left\langle E_{\alpha}\right|\hat{H}_{SB}\left|E_{\beta}\right\rangle \left\langle E_{\beta}\right|\hat{H}_{SB}\left|E_{\alpha}\right\rangle}{E_{\alpha}-E_{\beta}}+...,\label{eq:QDPT_H}
\eea
{\color{black} where higher order terms} in ${\left\langle E_{\beta}\right|\hat{H}_{SB}\left|E_{\alpha}\right\rangle}/{(E_{\alpha}-E_{\beta})}$ $(\beta\neq \alpha)$ are not shown explicitly.

{\color{black}
In the regime of large enough field and exchange interaction where ${\left|{\left\langle E_{\beta}\right|\hat{H}_{SB}\left|E_{\alpha}\right\rangle}/{(E_{\alpha}-E_{\beta})}\right|\ll1}$, the second row of Eq.~\ref{eq:QDPT_H} can be ignored. Furthermore, terms in $\left\langle E_{\alpha}\right|\hat{H}_{SB}\left|E_{\alpha}\right\rangle$ and in $\hat{H}_{B}$ that do not preserve the total Zeeman energy of the nuclei can be dropped in a secular approximation fashion. The resulting effective nuclear spin Hamiltonian $\hat{H}_{B;{\alpha}}$ reads
\bea
&&\hat{H}_{B;{\alpha}}=E_{\alpha}+\hat{H}_{BZ}+\hat{H}_{\alpha},\label{eq:H_Balpha_strong_B_J}
\eea
with $\hat{H}_{BZ}=-\gamma_{p}B\sum_{n}\hat{I}^z_{n}$  being the nuclear Zeeman energy, and the nuclear Hamiltonians $\hat{H}_{\alpha}$ includes secular hyperfine and nuclear spin flip-flop interactions,

\bea
\hat{H}_{S}&=&\sum_{n\neq m}d_{nm}\hat{I}_{n}^{+}\hat{I}_{m}^{-}+\sum_{n\neq m}(\text{\textminus}2d_{nm})\hat{I}_{nz}\hat{I}_{mz},\nonumber\\
\hat{H}_{-1}&=&-\frac{1}{2}\sum_{n}\left(A_{1nzz}+A_{2nzz}\right)\cdot \hat{I}_{nz}+\sum_{n\neq m}d_{nm}\hat{I}_{n}^+\hat{I}_{m}^-+\sum_{n\neq m}(\text{\textminus}2d_{nm})\hat{I}_{nz}\hat{I}_{mz},\nonumber\\
\hat{H}_{0}&=&\sum_{n\neq m}d_{nm}\hat{I}_{n}^+\hat{I}_{m}^-+\sum_{n\neq m}(\text{\textminus}2d_{nm})\hat{I}_{nz}\hat{I}_{mz},\nonumber\\
\hat{H}_{1}&=&\frac{1}{2}\sum_{n}\left(A_{1nzz}+A_{2nzz}\right)\cdot \hat{I}_{nz}+\sum_{n\neq m}d_{nm}\hat{I}_{n}^+\hat{I}_{m}^-+\sum_{n\neq m}(\text{\textminus}2d_{nm})\hat{I}_{nz}\hat{I}_{mz}.\label{eq:H_alpha_strong_B_J}
\eea
Here 
\beq
d_{nm}=-\frac{\mu_{0}}{8}\gamma_{p}^2\hbar\frac{1-3\cos^{2}\theta_{nm}}{R_{nm}^{3}},\label{eq:d_nm}
\eeq
is the dipolar interaction strength between protons $n$ and $m$, in which $R_{nm}$ is the length of the position vector $\vec{R}_{nm}$ joining $n$ and $m$, and $\theta_{nm}$ is the angle between $\vec{R}_{nm}$ and the field direction $z$. $A_{1nzz}$ ($A_{2nzz}$) is the $zz$ component of the hyperfine interaction tensor $\overleftrightarrow{A}_{1n}$ ($\overleftrightarrow{A}_{2n}$) between electron spin $1$ ($2$) and proton $n$. 
Eq.~\ref{eq:H_alpha_strong_B_J} is the total electron-nuclei Hamiltonian that the analysis in Sect.~\ref{subsec:PCA_theory} will be based on.
}

{\color{black}
\subsection{Pair Correlation Approximation for the regime of large field and exchange interaction}\label{subsec:PCA_theory}

Before presenting gCCE results, we perform some theoretical analysis of the regime of large field and exchange interaction that will be helpful for understanding the physics behind the results. In this subsection, we apply the theory of the Pair-Correlation Approximation\cite{yao2006theory,Liu_2007} (PCA) to our two-electron-spin model for the scenario of $N=1$. In Sect.~\ref{subsec:FID_theory}, expressions of $L^0_{\alpha\beta}(t)$ ($\alpha\neq\beta$) in the regime of large field and exchange interaction is derived for the scenario of $N=0$.

The theory of PCA captures the dynamics of nuclear spin pair flip-flop processes, which is usually the dominant source of nuclei-induced decoherence. It describes the dynamics of the nuclear spin on the time scale of electron spin decoherence as consisting of elementary excitations of nuclear spin pair flip-flop and maps all such excitations to independent pseudospin-$\frac{1}{2}$'s transitioning from down to up state. 
Specifically, in PCA one first picks any random pure initial state of the homonuclear spin-$\frac{1}{2}$ bath that is expressed as a direct product of the up and down states $\left|\mathcal{J}\right>=\otimes_{l}\left|j_{l}\right\rangle$,  where $\left|j_{l}\right\rangle=\left|\uparrow_l\right>$ or $\left|\downarrow_l\right>$ is an eigenstate of $\hat{I}_{lz}$ of nuclear spin $l$. 
The Hahn-echo decoherence profile of the electron spin system is independent of the random choice of $\left|\mathcal{J}\right>$ for large baths, as long as the up and down spins are approximately balanced in number and are both evenly distributed across the space. 
Each possible flip-flop transition from $\left|\mathcal{J}\right>=\left|\downarrow_{n}\uparrow_{m}\right>\otimes_{l\neq n,m} \left|j_{l}\right\rangle$ to $\left|\uparrow_{n}\downarrow_{m}\right>\otimes_{l\neq n,m} \left|j_{l}\right\rangle$ (shorthanded as $\left|\downarrow_{n}\uparrow_{m}\right>\rightarrow\left|\uparrow_{n}\downarrow_{m}\right>$)
is mapped to 
the transition of an independent pseudospin-$\frac{1}{2}$ labeled by $k$ from $\left|\downarrow_{k}\right>$ to 
$\left|\uparrow_{k}\right>$. 
Here $n$ ($m$) goes over any nuclear spin initially in the down (up) state, leading to a total $N_{\downarrow}\cdot N_{\uparrow}$ of independent pseudospins, with $N_{{\downarrow}/\uparrow}$ being the number of down/up spins in $\left|\mathcal{J}\right>$. 
The dynamics of the flip-flop transition is mapped to a rotation of the corresponding pseudospin $k$ under an effective field, starting from $\left|\downarrow_{k}\right>$  in its spin space.
The many-body wave function of the nuclei system is mapped to a direct product of pseudospin states, with the initial wavefunction $\left|\mathcal{J}\right>$ mapped to the initial state of the pseudospins $\left|\Phi\right>=\otimes_{k}\left|\downarrow\right\rangle _{k}$. The nuclear Hamiltonian conditioned on each electron spin level is mapped to a sum of effective Zeeman terms, with each term representing an effective field on a pseudospin. 
In the end, the decay of the Hahn-echo coherence functions, which are equal to inner products between nuclear system states under bifurcated evolution starting from the same state, are expressed with dynamics of pseudospins with simple forms.
For more details, we refer the readers to Ref.~[\onlinecite{yao2006theory,Liu_2007}].

We apply the same PCA procedure above to our two-spin system described by Eqs.~\ref{eq:pure_dephasing_total_H},\ref{eq:H_Balpha_strong_B_J},\ref{eq:H_alpha_strong_B_J} in the large field and exchange regime. Specifically, each nuclear Hamiltonian $\hat{H}_{\alpha}$ in Eq.~\ref{eq:H_alpha_strong_B_J} is mapped to effective fields on pseudospins $\sum_{k}\overrightarrow{\chi}_{\alpha k}\cdot\hat{\overrightarrow{\sigma}}_{k}/2$.
Normalized Hahn echo coherences can then be expressed as products of contributions from each nuclear pair flip-flop $\left|\downarrow_{n}\uparrow_{m}\right>\rightarrow\left|\uparrow_{n}\downarrow_{m}\right>$, which is mapped to rotations under effective fields of the corresponding pseudospin $k$.
In the following, we also refer to a nuclear spin pair flip-flop using the label $k$ of its corresponding pseudospin. 

\bea
L^{1}_{-1,0}(t=2\tau)&=&\prod_{k}\left|\left\langle \downarrow_{k}\right|e^{i\overrightarrow{\chi}_{0; k}\cdot\hat{\overrightarrow{\sigma}}_{k}\tau}e^{-i\overrightarrow{\chi}_{-1; k}\cdot\hat{\overrightarrow{\sigma}}_{k}\tau/2}e^{-i\overrightarrow{\chi}_{1; k}\cdot\hat{\overrightarrow{\sigma}}_{k}\tau/2}\left|\downarrow_{k} \right >\right|,\label{eq:L1_-1_0_PCA_k_resolution}\\
L^{1}_{-1,1}(2\tau)&=&\prod_{k}\left|\left\langle \downarrow_{k}\right|e^{i\overrightarrow{\chi}_{-1; k}\cdot\hat{\overrightarrow{\sigma}}_{k}\tau/2}e^{i\overrightarrow{\chi}_{1; k}\cdot\hat{\overrightarrow{\sigma}}_{k}\tau/2}e^{-i\overrightarrow{\chi}_{-1; k}\cdot\hat{\overrightarrow{\sigma}}_{k}\tau/2}e^{-i\overrightarrow{\chi}_{1; k}\cdot\hat{\overrightarrow{\sigma}}_{k}\tau/2}\left|\downarrow_{k}\right>\right|,\label{eq:L1_-1_1_PCA_k_resolution}\\
L^{1}_{-1,S}(2\tau)&=&\prod_{k}\left|\left\langle \downarrow_{k}\right|e^{i\overrightarrow{\chi}_{S; k}\cdot\hat{\overrightarrow{\sigma}}_{k}\tau}e^{-i\overrightarrow{\chi}_{-1; k}\cdot\hat{\overrightarrow{\sigma}}_{k}\tau/2}e^{-i\overrightarrow{\chi}_{1; k}\cdot\hat{\overrightarrow{\sigma}}_{k}\tau/2}\left|\downarrow_{k}\right>\right|,\label{eq:L1_-1_S_PCA_k_resolution}\\
L^{1}_{S,0}(2\tau)&=&\prod_{k}\left|\left\langle \downarrow_{k}\right|e^{i\overrightarrow{\chi}_{0; k}\cdot\hat{\overrightarrow{\sigma}}_{k}\tau}e^{-i\overrightarrow{\chi}_{S; k}\cdot\hat{\overrightarrow{\sigma}}_{k}\tau}\left|\downarrow_{k}\right>\right|=1,\\
L^{1}_{S,1}(2\tau)&=&\prod_{k}\left|\left\langle \downarrow_{k}\right|e^{i\overrightarrow{\chi}_{-1; k}\cdot\hat{\overrightarrow{\sigma}}_{k}\tau/2}e^{i\overrightarrow{\chi}_{1; k}\cdot\hat{\overrightarrow{\sigma}}_{k}\tau/2}e^{-i\overrightarrow{\chi}_{S; k}\cdot\hat{\overrightarrow{\sigma}}_{k}\tau}\left|\downarrow_{k}\right>\right|,\label{eq:L1_S_1_PCA_k_resolution}\\
L^{1}_{0,1}(2\tau)&=&\prod_{k}\left|\left\langle \downarrow_{k}\right|e^{i\overrightarrow{\chi}_{-1; k}\cdot\hat{\overrightarrow{\sigma}}_{k}\tau/2}e^{i\overrightarrow{\chi}_{1; k}\cdot\hat{\overrightarrow{\sigma}}_{k}\tau/2}e^{-i\overrightarrow{\chi}_{0; k}\cdot\hat{\overrightarrow{\sigma}}_{k}\tau}\left|\downarrow_{k}\right>\right|.\label{eq:L1_0_1_PCA_k_resolution}
\eea
Here $\hat{\overrightarrow{\sigma}}_{k}$ is the Pauli matrix vector for pseudospin $k$ and the effective fields have cartesian components
\beq
\overrightarrow{\chi}_{\alpha; k}=(2C_{\alpha k},0,D_{\alpha k}+E_{\alpha k}),
\eeq
where the flip-flop transition matrix element is equal to the dipolar interaction as in Eq.~\ref{eq:d_nm}
\beq
C_{\alpha k}=d_{nm},
\eeq
and the energy costs of the flip-flop $D_{\alpha k}+E_{\alpha k}$ are
\bea
D_{\alpha k}
&=&-4\sum_{m'\neq n,m}(d_{ nm'}-d_{ mm'})j_{m'}\label{eq:D_alpha_k}\\
E_{Sk}&=&0\\
E_{0k}&=&0\\
E_{-1k}=-E_{1k}&=&-\frac{1}{2}[\left(A_{1nzz}+A_{2nzz}\right)-\left(A_{1mzz}+A_{2mzz}\right)],
\eea
with $j_{m'}={\frac{1}{2}}$ ($j_{m'}=-\frac{1}{2}$) for up (down) spins in the initial nuclear state $\left|\mathcal{J}\right>$, and $A_{inzz}$ is the $zz$ component of the dipolar hyperfine interaction as in Eq~\ref{eq:dipolar_hyperfine}.
The order of time evolution under different effective fields in Eqs.~\ref{eq:L1_-1_0_PCA_k_resolution}-\ref{eq:L1_0_1_PCA_k_resolution} reflects the behavior of the $\pi_{xx}$ pulse that flips $\left|E_{-1}\right\rangle$ and $\left|E_{1}\right\rangle$ while keeping $\left|E_{0}\right\rangle$ and $\left|E_{S}\right\rangle$ unchanged.

Each pair contribution is an overlap (inner product) between two pseudospin states traversing different pathways from the same state $\left|\downarrow_{k}\right>$ in the spin space under different effective fields. 
A sketch of the effective fields $\overrightarrow{\chi}_{\alpha k}$ is shown in Fig.~\ref{fig:sketch_of_kis}. For example, the contribution to $L^{1}_{-1,0}$ from a pair flip flop $k$, $\left|\left\langle \downarrow_{k}\right|e^{i\overrightarrow{\chi}_{0; k}\cdot\hat{\overrightarrow{\sigma}}_{k}\tau}e^{-i\overrightarrow{\chi}_{-1; k}\cdot\hat{\overrightarrow{\sigma}}_{k}\tau/2}e^{-i\overrightarrow{\chi}_{1; k}\cdot\hat{\overrightarrow{\sigma}}_{k}\tau/2}\left|\downarrow_{k} \right >\right|\equiv\left|\left\langle \psi_1\right|\psi_2\right>|$, where $\left|\psi_1\right>$ is the result of  $\left|\downarrow_{k}\right>$ precessing about the field $\overrightarrow{\chi}_{0,k}$ for $2\tau$ time, and $\left|\psi_2\right>$ is $\left|\downarrow_{k}\right>$ precessing about the field $\overrightarrow{\chi}_{1,k}$ for $\tau$ time and then about the field $\overrightarrow{\chi}_{-1,k}$ for the second time $\tau$ in the Hahn echo sequence. 

\begin{figure}[htp]
$\begin{array}{c}
\includegraphics[width=0.4\columnwidth]{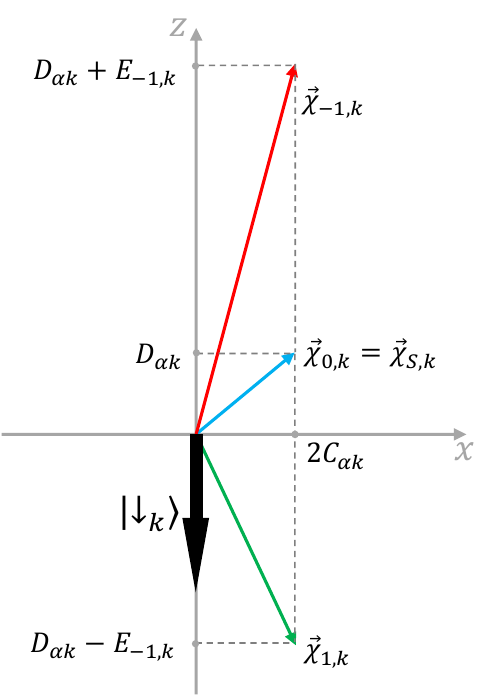}
\end{array}$\caption{{\color{black} A sketch of the effective fields $\overrightarrow{\chi}_{\alpha k}$ on the pseudospins. Their cartesian components are also labeled on axis. The black arrow represents the Bloch vector of $\left|\downarrow_{k}\right>$.
}}\label{fig:sketch_of_kis}
\end{figure}

Since $\overrightarrow{\chi}_{0; k}=\overrightarrow{\chi}_{S; k}$, at strong field and exchange, the PCA predicts no decoherence in $L^{1}_{S,0}$. This is consistent with the well-known fact that singlet-triplet qubits with computational basis $\left|E_S\right\rangle$ and $\left|E_0\right\rangle$ are insensitive to the magnetic fluctuation to the leading order\cite{wang2012composite,wu2014two}. The decoherence between $\left|E_S\right\rangle$ and $\left|E_0\right\rangle$ is due to higher-order terms in Eq.~\ref{eq:QDPT_H}\cite{zhang2020cluster} which are omitted in the large field and exchange regime.
Also following from $\overrightarrow{\chi}_{0; k}=\overrightarrow{\chi}_{S; k}$ are $L^{1}_{-1,0}=L^{1}_{-1,S}$ and $L^{1}_{S,1}=L^{1}_{0,1}$. 
With the Baker-Campbell-Hausdorff formula $e^{\hat{A}}e^{\hat{B}}\approx e^{\hat{A}+\hat{B}+\frac{1}{2}\left[\hat{A},\hat{B}\right]}$, one can also show that $L^{1}_{-1,0}\approx   \prod_{k}\left|\left\langle \downarrow_{k}\right|e^{-\frac{\tau^{2}}{8}\left[\overrightarrow{\chi}_{-1;k}\cdot\hat{\overrightarrow{\sigma}}_{k},\overrightarrow{\chi}_{1;k}\cdot\hat{\overrightarrow{\sigma}}_{k}\right]}\left|\downarrow_{k}\right>\right| \approx L^{1}_{0,1}$. Therefore, the four off-diagonal coherences $L^{1}_{-1,0},L^{1}_{-1,S},L^{1}_{S,1},L^{1}_{0,1}$ are approximately the same at large field and exchange within PCA.

Here we make some analysis of the pair contribution for $L^{1}_{-1,0}$ and $L^{1}_{-1,1}$ that will be crucial for our understanding of dependence of Hahn-echo coherence time on various model parameters. For $L^{1}_{-1,0}$, let 
\beq
f_k\equiv 1-\left|\left\langle \downarrow_{k}\right|e^{i\overrightarrow{\chi}_{0; k}\cdot\hat{\overrightarrow{\sigma}}_{k}\tau}e^{-i\overrightarrow{\chi}_{-1; k}\cdot\hat{\overrightarrow{\sigma}}_{k}\tau/2}e^{-i\overrightarrow{\chi}_{1; k}\cdot\hat{\overrightarrow{\sigma}}_{k}\tau/2}\left|\downarrow_{k} \right >\right|,\label{eq:f_k}
\eeq
be a measure of decoherence due to pair $k$. $f_k$ is a function of $C_{\alpha k}$($=d_{nm}$), $D_{\alpha k}$, $E_{-1,k}$ and $\tau$. Inspecting the precessions of $\left|\downarrow_{k} \right >$ under the effective fields in Fig.~\ref{fig:sketch_of_kis}, it is not hard to see that for any $D_{\alpha k}$ and $\tau$, $f_k(C_{\alpha k}=0,E_{-1,k})=0$ and $f_k(C_{\alpha k},E_{-1,k}=0)=0$. Since $f_k$ is a non-negative smooth function of $C_{\alpha k}$ and $E_{-1,k}$, it indicates that $\frac{\partial f_k}{\partial C_{\alpha k}}|_{C_{\alpha k}=0}=0$, $\frac{\partial^2 f_k}{\partial C_{\alpha k}^2}|_{C_{\alpha k}=0}>0$ and $\frac{\partial f_k}{\partial E_{-1, k}}|_{E_{-1, k}=0}=0$, $\frac{\partial^2 f_k}{\partial E_{-1, k}^2}|_{E_{-1, k}=0}>0$, which means, at least for small $|C_{\alpha k}|$ and $|E_{-1,k}|$, a nuclear spin pair $k$ needs larger $|C_{\alpha k}|$ and $|E_{-1,k}|$ to contribute to greater decoherence. Following the same line of reasoning and considering the decoherence contributed by nuclear spin pairs to $L^{1}_{-1,1}$ defined as 
\beq
g_k\equiv1-\left|\left\langle \downarrow_{k}\right|e^{i\overrightarrow{\chi}_{-1; k}\cdot\hat{\overrightarrow{\sigma}}_{k}\tau/2}e^{i\overrightarrow{\chi}_{1; k}\cdot\hat{\overrightarrow{\sigma}}_{k}\tau/2}e^{-i\overrightarrow{\chi}_{-1; k}\cdot\hat{\overrightarrow{\sigma}}_{k}\tau/2}e^{-i\overrightarrow{\chi}_{1; k}\cdot\hat{\overrightarrow{\sigma}}_{k}\tau/2}\left|\downarrow_{k}\right>\right|,\label{eq:g_k}
\eeq
the same conclusion can be drawn. 

Both $f_k$ and $g_k$ are relatively insensitive to the typical value of $D_{\alpha k}$ that is on the order of the dipolar interaction $d_{nm}$.
We plot pair-contributed decoherence $f_k$ and $g_k$ as functions of $C_{\alpha k}$ and $E_{-1,k}$ in Fig.~\ref{fig:fk_gk_vs_C_E}. The properties of the above-mentioned partial derivatives are confirmed in the figure. Both $f_k$ and $g_k$ increase monotonically with $\left|C_{\alpha k}\right|$. 
For $f_k$ to be large, it generally requires large enough $|E_{-1, k}|$. For $g_k$ to be large, $|E_{-1,k}|$ must be large enough as well, and there is an optimum range of $|E_{-1,k}|$ that maximizes $g_k$. The vast majority of nuclear spin pairs reside near the center of these plots within the region of parabolic dependence on $C_{\alpha k}$ and $E_{-1,k}$, contributing little to electron decoherence.  

\begin{figure}[htp]
$\begin{array}{c}
\includegraphics[width=0.9\columnwidth]{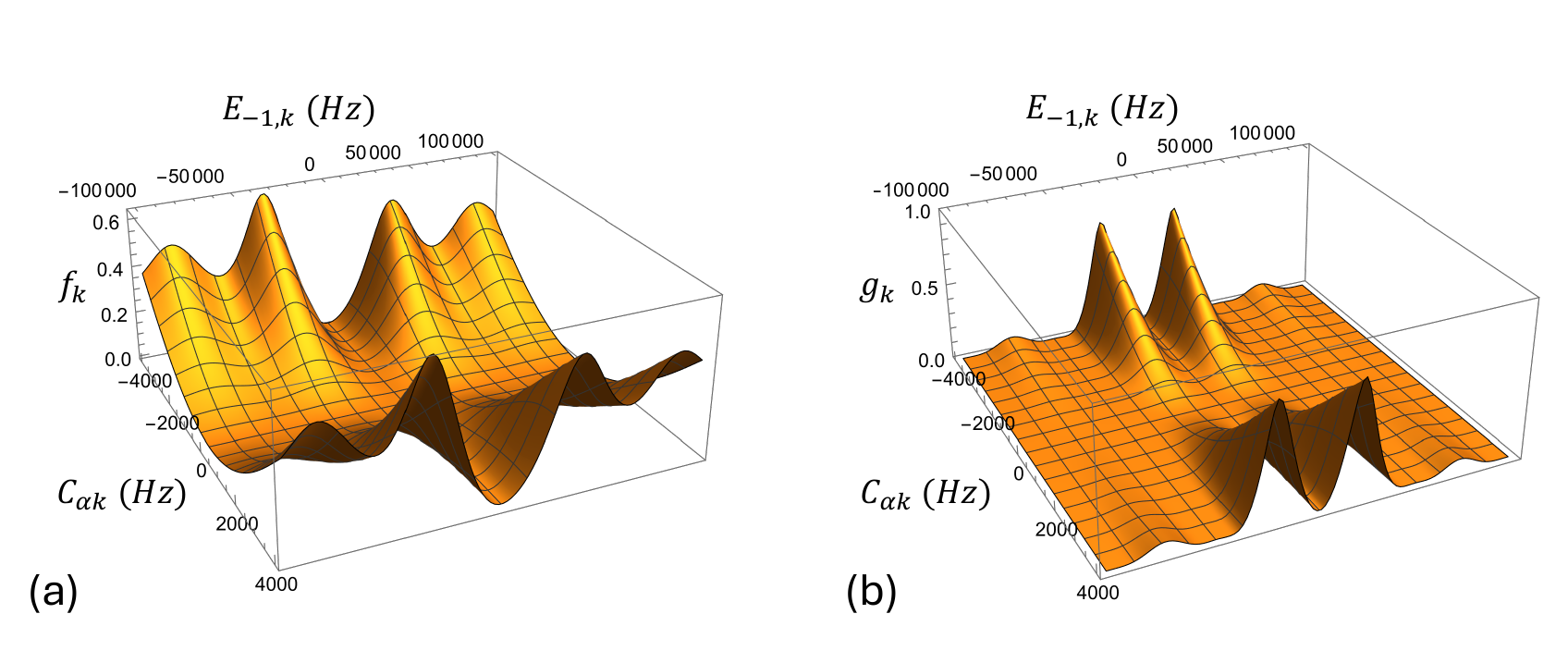}
\end{array}$\caption{{\color{black} 
(a) $f_k$ and (b) $g_k$ as functions of $C_{\alpha k}$ and $E_{-1,k}$. For these plots, $D_{\alpha k}=5\textrm{kHz}$, $\tau=20\mus$.
}}\label{fig:fk_gk_vs_C_E}
\end{figure}

Therefore, pair flip-flops that contribute noticeably to decoherence in $L^{1}_{-1,0}$ (and $L^{1}_{-1,S},L^{1}_{S,1},L^{1}_{0,1}$) and $L^{1}_{-1,1}$ 
are associated with large dipolar interaction $\left|d_{nm}\right|$, which controls the pair flip-flop rate 
(Eq.~\ref{eq:H_alpha_strong_B_J}), 
as well as large enough $\left|E_{-1,k}\right|$, which is the difference in total secular hyperfine interaction of each proton spin of a pair. 
$\left|E_{-1,k}\right|$ describes the difference in the dipolar field on the electron spin system from each spin in a pair and is a measure of the amplitude of fluctuating magnetic field due to the pair flip-flop process. 
The following properties are in turn satisfied by
these pair flip-flops that cause decoherence: (1) according to Eq.~\ref{eq:d_nm}, the position vector $\vec{R}_{nm}$ joining the two protons is short and is either near parallel or near perpendicular to the external field. (2) Since $\vec{R}_{nm}$ is short, $\left|E_{-1,k}\right|=\frac{1}{2}|\left(A_{1mzz}-A_{1nzz}\right)+\left(A_{2mzz}-A_{2nzz}\right)|\simeq\frac{1}{2}|(\nabla_{\vec{r}} A_{1zz}+\nabla_{\vec{r}} A_{2zz})|_{\vec{r}=\vec{R_c}}\cdot \vec{R}_{nm}|$, where $A_{1zz}$($A_{2zz}$) is the secular hyperfine interaction $A_{1nzz}$($A_{2nzz}$) as a scalar function of the position $\vec{r}$ of the proton $n$. The field $\nabla_{\vec{r}}( A_{1zz}+A_{2zz})|_{\vec{r}=\vec{R}}$ at the location of the center $\vec{R_c}$ of the pair is large enough and nearly parallel to $\vec{R}_{nm}$. 
In Sect.~\ref{sec:results_and_discussion}, these two properties will help us gain physical insight into the dependence of the Hahn-echo coherence times on various model parameters.

\subsection{FID ($N=0$) in the regime of large field and exchange interaction}\label{subsec:FID_theory}

In this section, we derive the expression for $T_2^*$ for the scenario of $N=0$ in the regime of large field and exchange. Again, we start with the projected Hamiltonian in Eqs.~\ref{eq:pure_dephasing_total_H},\ref{eq:H_Balpha_strong_B_J},\ref{eq:H_alpha_strong_B_J}. The dephasing in the free induction decay is expected to be due to inhomogeneous dephasing, which for our model is a dephasing from averaging electron coherence over different Overhauser fields corresponding to different pure states of the nuclei. The state of the nuclei can be viewed as frozen on the time scale of dephasing, and only the coupling between electron spins to each single proton needs to be kept, whereas all the proton-proton interactions can be dropped from the Hamiltonian. The dephasing in the $N=0$ scenario being a single proton effect is further confirmed by gCCE even in smaller fields and exchange (see Sect.~\ref{sec:results_and_discussion}). Therefore, we use the projected Hamiltonian Eqs.~\ref{eq:pure_dephasing_total_H},\ref{eq:H_Balpha_strong_B_J} with $\hat{H}_{\alpha}$ of
\bea
\hat{H}_{S}&=&0\nonumber\\
\hat{H}_{-1}&=&-\frac{1}{2}\sum_{n}\left(A_{1nzz}+A_{2nzz}\right)\cdot \hat{I}_{nz}\nonumber\\
\hat{H}_{0}&=&0\nonumber\\
\hat{H}_{1}&=&\frac{1}{2}\sum_{n}\left(A_{1nzz}+A_{2nzz}\right)\cdot \hat{I}_{nz}\label{eq:H_alpha_strong_B_J_FID}
\eea
Initial state of the full system reads,
\beq
\hat{\rho}(0)=\hat{\rho}_{S}(0)\otimes\sum_{\mathcal{J}}P_{\mathcal{J}}|\mathcal{J}><\mathcal{J}|,
\eeq
where we have rewritten $\hat{\rho}_B(0)=\mathbb{I}_{2^{N_B}\times 2^{N_B}}$ as a classical superposition of all pure states of the form $\left|\mathcal{J}\right\rangle=\otimes_{n}\left|j_{n}\right\rangle$ ($\left|j_{n}\right\rangle=\left|\uparrow_n\right>$ or $\left|\downarrow_n\right>$) with equal probability $P_{\mathcal{J}}$.
A nuclear spin state of the form $\left|\mathcal{J}\right\rangle$ is a common eigenstate of Eqs.~\ref{eq:H_alpha_strong_B_J_FID}, with the eigenvalues serving as Overhauser fields acting on the electron spin states $\left|E_{\alpha}\right\rangle$.
Given the time evolution of the density matrix of the entire system for $N=0$,
\beq
\hat{\rho}_{S}(t)=\mathrm{Tr}_{B}\left[e^{-iH_{proj}t}\hat{\rho}(0)e^{iH_{proj}t}\right],
\eeq
One can show,
\beq
L^0_{\alpha\beta}=\left|\sum_{\mathcal{J}}P_{\mathcal{J}}<\mathcal{J}_{\beta}|\mathcal{J}_{\alpha}>\right|\label{eq:L0_alpha_beta}
\eeq
where
\beq
|\mathcal{J}_{\alpha}>=e^{-i\hat{H}_{\alpha}t}|\mathcal{J}>.
\eeq
Consequently,
\bea
&&L^0_{-1,1}(t)=L^0_{-1,S}(2t)=L^0_{-1,0}(2t)=L^0_{1,S}(2t)=L^0_{1,0}(2t)\nonumber\\
&&=\left|\sum_{\mathcal{J}}P_{\mathcal{J}}<\mathcal{J}_{1}|\mathcal{J}_{-1}>\right|=\left|\sum_{\mathcal{J}}P_{\mathcal{J}}e^{i\sum_{n}\left(A_{1nzz}+A_{2nzz}\right)\cdot j_{n}t}\right|\label{eq:L0_-1_1}
\eea
\bea
L^0_{S,0}(t)=1
\eea
Here we see that $L^0_{-1,S}$, $L^0_{-1,0}$, $L^0_{1,S}$ and $L^0_{1,0}$ share the same dephasing profile, while $L^0_{S,0}$ decays twice as fast as them. $L^0_{S,0}$ shows no dephasing in the limit of the strong field and exchange.
Consider the quantity inside the modulus sign in Eq.~\ref{eq:L0_-1_1}
\beq
\sum_{\mathcal{J}}P_{\mathcal{J}}e^{i\sum_{n}\left(A_{1nzz}+A_{2nzz}\right)\cdot j_{n}t}=\sum_{\mathcal{J}}P_{\mathcal{J}}e^{-iE_{\mathcal{J}}t}=\int P(E)e^{-iEt}dE,
\eeq
where the hyperfine energy due to the Overhauser field of a certain nuclei
state is $E_{\mathcal{J}}=\sum_{n}\left(A_{1nzz}+A_{2nzz}\right)\cdot(-j_{n})$,
and its inhomogeneous broadening distribution is defined as
\beq
P(E)=\sum_{\mathcal{J}}P_{\mathcal{J}}\delta(E-E_{\mathcal{J}}).
\eeq
For large total number of nuclei spins, this distribution approaches
normal distribution due to the Lindeberg–Feller 
 Central Limit Theorem
\beq
P(E)=\frac{1}{\sqrt{2\pi\sigma^{2}}}e^{-\frac{E^{2}}{2\sigma^{2}}},
\eeq
where the variance
\beq
\sigma^{2}=\sum_{n}\frac{1}{4}\left(A_{1nzz}+A_{2nzz}\right)^{2}.\label{eq:sigma_variance}
\eeq
Therefore,
\beq
L^0_{-1,1}=\int P(E)e^{-iEt}dE=e^{-(t/T_{2}^{*})^{2}},\label{eq:L0_-1,1_form}
\eeq
where
\beq
T_{2}^{*}=\sqrt{2}/\sigma.\label{eq:T2_star_theory}
\eeq

}

\section{Results and discussion}\label{sec:results_and_discussion}

{\color{black} In this section, we present gCCE results of the decoherence of the two electrons coupled to a nuclear spin bath as described in Sect.~\ref{subsec:model}.}
Because we are in the pure dephasing regime, the diagonal elements of the two-electron RDM are assumed to be constant and we focus on extracting coherence times from off-diagonal $L^N_{\alpha\beta}(t)$ for $\alpha\neq\beta$ by fitting with a stretched exponential $\exp\left[-(t/T^N_{\alpha\beta})^b\right]$. The primary results are the coherence times when the system undergoes FID and upon applying a Hahn echo sequence, i.e. $T^0_{\alpha\beta}\equiv T_{2;\alpha\beta}^*$ and $T^1_{\alpha\beta}\equiv T_{2;\alpha\beta}$, respectively, as functions of the system parameters $B$, $J$, $\vec{d}$, $R_S$, etc., in the ranges typical for molecular magnetic systems described in Sect.~\ref{subsec:model}.
{\color{black} Physical understanding of the parameter dependence of coherence times with the help of the theoretical considerations in Sects.~\ref{subsec:PCA_theory} and \ref{subsec:FID_theory} are also presented.
Implications for experiments on achieving longer coherence times in molecular two-spin-$\frac{1}{2}$ systems are provided in Sect.~\ref{subsec:implications}.
}

Before proceeding to the main results, we comment on the convergence of the gCCE used to calculate $L^N_{\alpha\beta}$.  For FID (Hahn echo), {\color{black} on the time scale of decoherence of $L^{0(1)}_{-1,0}, L^{0(1)}_{1,0}, L^{0(1)}_{-1,S},L^{0(1)}_{1,S}$ and $L^{0(1)}_{-1,1}$ }, the cluster correlation expansion converges well at gCCE order 1 (order 2) as exemplified in the Supplementary Material  Figs.~S1 and S2, i.e. $T_2^*$ ($T_2$) is dominated by irreducible correlations from single (pairs of) nuclear spins. This is a result of the slow, as compared to the other time scales in the Hamiltonian, nuclear dynamics. It indicates that the $T_2^*$ mechanism is inhomogeneous dephasing from averaging the coherence functions over random total Overhauser fields from all possible pure initial nuclear bath states.\cite{yao2006theory,dobrovitski2008decoherence} This averaging process is implicitly included when we use the complete mixed state for the initial bath. Convergence at gCCE order 2 in the Hahn-echo scenario indicates that the $T_2$ processes are dominated by the nuclear spin pair flip-flop processes\cite{yang2008quantum,yang2009quantum}.

\begin{figure}[htp]
$\begin{array}{cc}
\includegraphics[width=0.47\columnwidth]{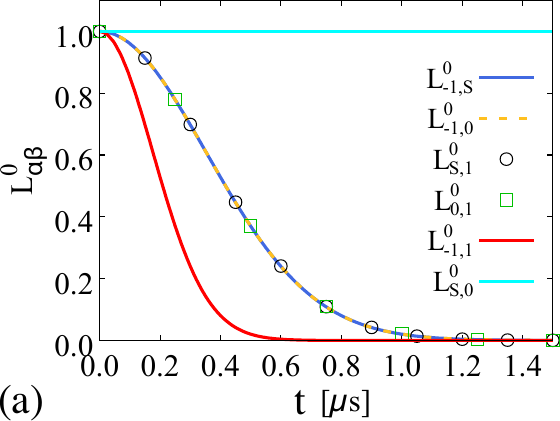}
~~~~~
\includegraphics[width=0.47\columnwidth]{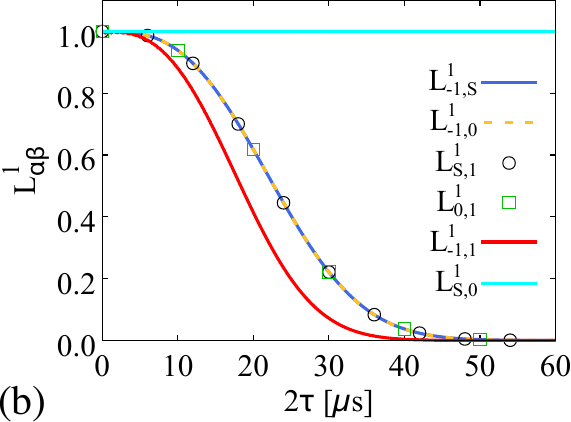}
\end{array}
$\caption{(a) Time evolution of coherences $L^0_{\alpha\beta}$, in the scenario of FID ($N=0$); (b) Time evolution of coherences $L^1_{\alpha\beta}$, in the scenario of Hahn-echo ($N=1$).
In both cases $N=0$ and $N=1$, $L^N_{-1,0}=L^N_{1,0}=L^N_{-1,S}=L^N_{1,S}\neq L^N_{-1,1}$, and these coherences exhibit much faster decay in contrast to $L^N_{S,0}$, which exhibits essentially no change on the decay time scale of other coherences.
In the example plotted here, the model parameters are $J=10\GHz$, $d=5\ang$, $R_S=5\ang$, $R_B=2\ang$, $n_B=0.01/\ang^{3}$, $B=1\T$, and two electrons are aligned along $z$, i.e. parallel to the field.
}\label{fig:three_groups_RDM}
\end{figure}

{\color{black}
\subsection{gCCE results of coherence times}\label{subsec:gCCE results}

Within the range of $B$ and $J$ we consider, in the gCCE results it is observed that $L^N_{-1,0}=L^N_{1,0}=L^N_{-1,S}=L^N_{1,S}$ (Fig.~\ref{fig:three_groups_RDM}). The proof of this equality for both $N=1$ and $N=0$ in the large $B$ and $J$ regime was presented in Sects.~\ref{subsec:PCA_theory} and \ref{subsec:FID_theory}, which relies on the symmetrical form of the projected Hamiltonians in Eq.~\ref{eq:H_alpha_strong_B_J} and in Eq.~\ref{eq:H_alpha_strong_B_J_FID}. 
The symmetrical form of more general projected Hamiltonians, such as Eq.~\ref{eq:QDPT_H} which includes higher-order terms and non-secular nuclear terms, leads to this equality between off-diagonal coherences at smaller $B$ and $J$.
$L^N_{-1,1}$ is in general different from the above four coherences. This difference for the $N=1$ scenario originates from the difference within the four projected nuclear Hamiltonians and that the $\pi_{xx}$ pulse affects the $|E_\alpha\rangle$'s differently, leading to a significantly different evolution path of the pseudospin states as in Eq.~\ref{eq:L1_-1_1_PCA_k_resolution} in contrast to Eq.~\ref{eq:L1_-1_0_PCA_k_resolution},\ref{eq:L1_-1_S_PCA_k_resolution},\ref{eq:L1_S_1_PCA_k_resolution} and \ref{eq:L1_0_1_PCA_k_resolution}.
For the $N=0$ scenario, $L^0_{-1,1}$ decays twice as fast as $L^0_{-1,0}=L^0_{1,0}=L^0_{-1,S}=L^0_{1,S}$ in the regime of large $B$ and $J$, as shown in Eq.~\ref{eq:L0_-1_1}. This is because the nuclear states $|\mathcal{J}_{-1}\rangle$ and $|\mathcal{J}_{1}\rangle$ in Eq.~\ref{eq:L0_alpha_beta} evolve under hyperfine Hamiltonians with different sign, whereas $|\mathcal{J}_{0}\rangle$ and $|\mathcal{J}_{S}\rangle$ remain constant. This effectively causes a doubled Overhauser field on the two-level system spanned by $|E_{-1}\rangle$ and $|E_{1}\rangle$ compared to the two levels corresponding to $L^0_{-1,0},L^0_{1,0},L^0_{-1,S}$ and $L^0_{1,S}$.
There is little nuclear-spin-induced decoherence between $|E_S\rangle$ and $|E_0\rangle$ on the time scale of decoherence of other $L^N_{\alpha\beta}$ (Fig.~\ref{fig:three_groups_RDM}). For large $B$ and $J$, this is the result of a zero hyperfine coupling in both $\hat{H}_S$ and $\hat{H}_0$ in Eqs.~\ref{eq:H_alpha_strong_B_J} and \ref{eq:H_alpha_strong_B_J_FID}.

In this work, we are interested in the initial decay of the RDM as a whole entity, which is most relevant in experimental situations such as the decaying fidelity of two-qubit entangling gates and the measurement of the transverse magnetization decay in standard pulse EPR experiments. Therefore, we focus on the time scale of the decay of the five coherences except $L^N_{0,S}$. The decoherence of $L^N_{0,S}$ occurs on much longer time scales and has been an important subject of study for the realization of singlet-triplet qubits.\cite{wang2012composite,wu2014two}} 
To avoid redundant or trivial results, we plot the coherence times extracted from $L^N_{-1,S}$ and $L^N_{-1,1}$.

\begin{figure}[htp]
$\begin{array}{c}
\includegraphics[width=0.48\columnwidth]{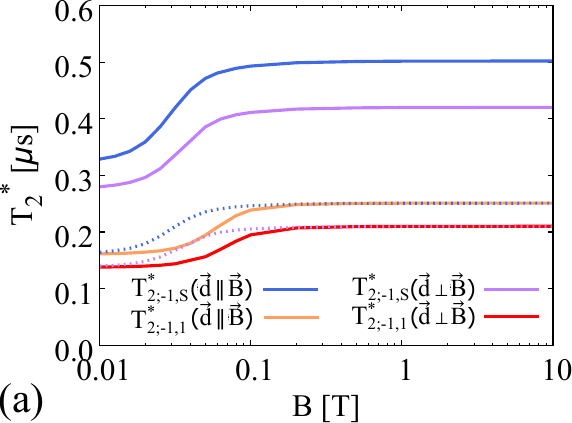}\\
\\
\includegraphics[width=0.46\columnwidth]{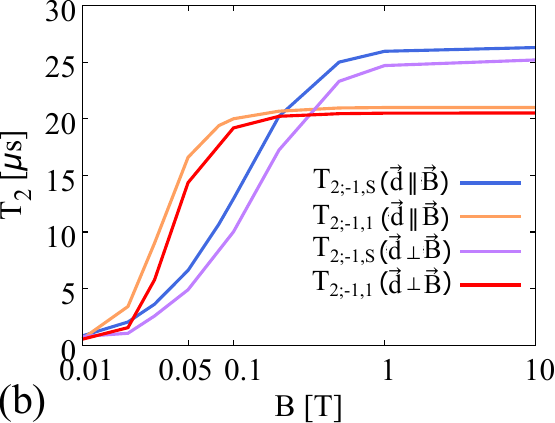}\\
\\
\includegraphics[width=0.48\columnwidth]{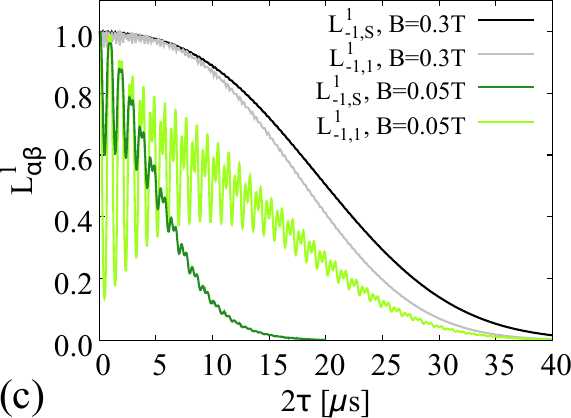}
\end{array}$\caption{Dependence of (a) $T_2^*$ and (b) $T_2$ on the field strength $B$ and relative orientation between the $\vec{B}$ field  and the position vector $\vec{d}$ joining the two electron spins.  
The dashed blue and purple lines in (a) are $0.5T^*_{2;-1,S}$ for $\vec{B}\parallel\vec{d}$ and $\vec{B}\perp\vec{d}$, respectively.
In (c), the time evolution of $L^1_{-1,S}$ and $L^1_{-1,1}$ at $B=0.3\T$ and $B=0.05\T$ for $\vec{B}\parallel\vec{d}$ are shown with ESEEM developed significantly at the smaller field between the two.
Other parameters of the model used for the results in this figure are $J=10\GHz$, $d=5\ang$, $R_S=5\ang$, $R_B=2\ang$, $n_B=0.01/\ang^{3}$.
}\label{fig:T2Star_vs_Bz_and_Bz_for_dx_vs_Bz}
\end{figure}

First, we consider the dependence of coherence times $T_2^*$ and $T_2$ on the strength $B$ of the external field and the relative orientation between the field $\vec{B}$ and the position vector $\vec{d}$ that joins the two spins of electrons. As shown in Figs.~\ref{fig:T2Star_vs_Bz_and_Bz_for_dx_vs_Bz}(a) and (b), at all field strengths of interest in the pure dephasing regime, and for both $L^N_{-1,S}$ and $L^N_{-1,1}$, $T_2^*$ and $T_2$ are larger when $\vec{B}\parallel\vec{d}$ as compared to $\vec{B}\perp\vec{d}$. {\color{black} This dependence of coherence times on relative orientation is most prominent when $\vec{d}$ is comparable to $2R_S$.}
Both the values of $T_2^*$ and $T_2$ remain constant at large enough fields{\color{black}, which corresponds to the regime of large fields in Sect.~\ref{subsec:PCA_theory} and \ref{subsec:FID_theory} where the coherence times are independent of $B$,} and start to decrease only at smaller fields. The ratio of $T^*_{2;-1,S}$ to that of $T^*_{2;-1,1}$ is two at the large fields{\color{black}, in agreement with Eq.~\ref{eq:L0_-1_1},} where each $T_2^*$ remains constant and becomes larger than two when the field decreases and $T_2^*$ starts to be sensitive to the field. Note that the dashed blue and purple lines in Fig.~\ref{fig:T2Star_vs_Bz_and_Bz_for_dx_vs_Bz}(a) are $0.5T^*_{2;-1,S}$ for $\vec{B}\parallel\vec{d}$ and $\vec{B}\perp\vec{d}$, respectively.
For the case of Hahn echo $T_2$, the decoherence of $L^1_{-1,S}$ is slower than $L^1_{-1,1}$ at large fields and becomes faster when the field decreases, as shown in Fig.~\ref{fig:T2Star_vs_Bz_and_Bz_for_dx_vs_Bz}(b). Oscillatory features in both $L^1_{-1,S}$ and $L^1_{-1,1}$ similar to electron spin echo envolope modulation (ESEEM)\cite{rowan1965electron} in the context of single electron spin echo develop at very small field as shown in Fig.~\ref{fig:T2Star_vs_Bz_and_Bz_for_dx_vs_Bz}(c). For the cases where the ESEEM feature is significant, $T_2$ was extracted by fitting to the upper envelope of the coherence.
{\color{black} The appearance of the ESEEM signifies that non-secular hyperfine interaction terms involving $I_{nx}$ and $I_{ny}$ operators that cause nuclear state transitions play a role\cite{schweiger2001principles}. Therefore, the ESEEM appears outside of the large field regime described by Eq.~\ref{eq:H_alpha_strong_B_J}, and the inclusion of non-secular hyperfine terms is required for an accurate theoretical description of small fields.}
In the following, we focus on large fields and the situation of $\vec{B}\parallel\vec{d}$ and study the effect of other model parameters on the coherence times, since $\vec{B}\parallel\vec{d}$ results in longer coherence times compared to $\vec{B}\perp\vec{d}$.
{\color{black}
The physics of this dependence of coherence times on relative field orientation will be analyzed in detail in Sect.~\ref{subsec:understanding}.
}

\begin{figure}[htp]
$\begin{array}{cc}
\includegraphics[width=0.4\columnwidth]{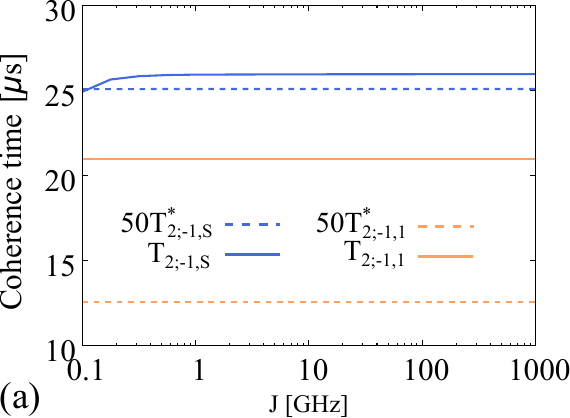}
~~~~~~~
\includegraphics[width=0.4\columnwidth]{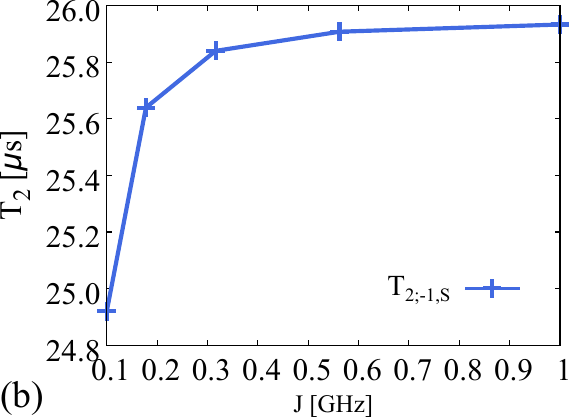}
\end{array}$\caption{(a) Dependence of $T_2$ and $T_2^*$ on the exchange interaction $J$ between the two electron spins. 
Solid (dashed) lines are $T_2$ ($50T_2^*$), and blue (orange) lines are the coherence times for $L^N_{-1,S}$ ($L^N_{-1,1}$).
Note that $T_2^*$ have been scaled up by a factor of 50. 
The same line and color style are used to label the coherence times in the following Figs.~\ref{fig:T2Star_T2_vs_dz}, \ref{fig:T2Star_T2_vs_r}, \ref{fig:T2Star_T2_vs_np}, \ref{fig:T2Star_T2_vs_s}.
{\color{black} (b) $T_{2;-1,S}$ at small J is plotted for showing the variation.}
Other parameters of the model used for the results in this figure are $d=5\ang$, $R_S=5\ang$, $R_B=2\ang$, $n_B=0.01/\ang^{3}$, $B=1\T$, $\vec{d}\parallel\vec{B}$.
}\label{fig:T2Star_T2_vs_J}
\end{figure}

The effect of tuning the magnitude of the exchange interaction $J$ between the two electron spins while keeping other parameters constant is shown in Fig.~\ref{fig:T2Star_T2_vs_J}{\color{black} (a)}.
Note that in Fig.~\ref{fig:T2Star_T2_vs_J} and the following figures we have multiplied $T_2^*$ by a factor of 50 and plot it in the same scale as $T_2$ to show an order of magnitude difference between the two and retain a linear vertical axis for coherence times to better exhibit small structures in curves.
{\color{black}
Any correlation between the dependence of $T_2^*$ and of $T_2$ on a model parameter is also better seen in this way. 
}
Essentially, in the {\color{black} wide} range of possible values of $J$ of interest, which crosses many orders of magnitude, both $T_2^*$ and $T_2$ are insensitive to $J$. The first exception occurs in $T_{2;-1,S}$, where the decoherence starts to become faster when $J$ is decreased towards the small limit {\color{black} (Fig.~\ref{fig:T2Star_T2_vs_J}(b))}.
{\color{black}
The insensitivity of coherence times to large enough $J$ is a result of the fact that the projected Hamiltonians Eqs.~\ref{eq:H_alpha_strong_B_J} and \ref{eq:H_alpha_strong_B_J_FID} which the nuclear spin system effectively experiences are independent of $J$. 
}

\begin{figure}[htp]
$\begin{array}{cc}
\includegraphics[width=0.48\columnwidth]{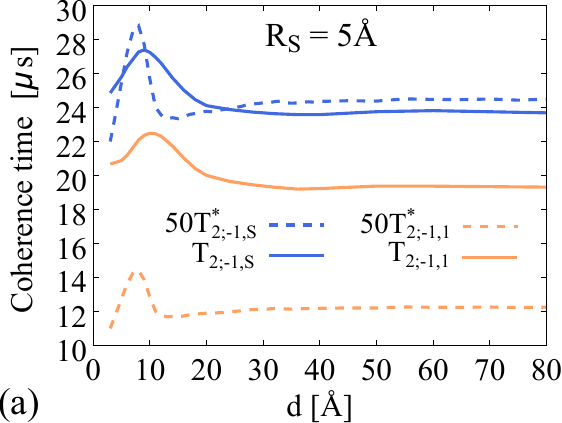}
~~~
\includegraphics[width=0.48\columnwidth]{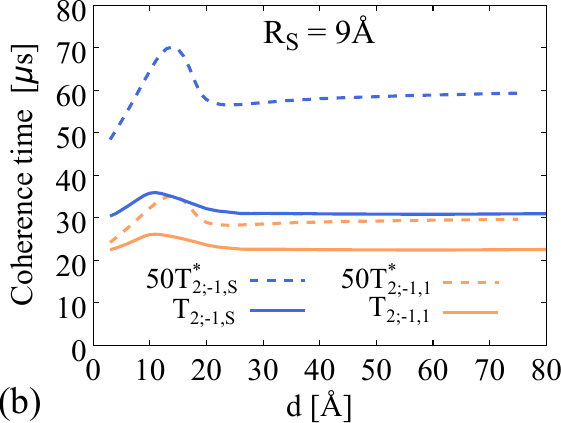}\\
\\
\includegraphics[width=0.48\columnwidth]{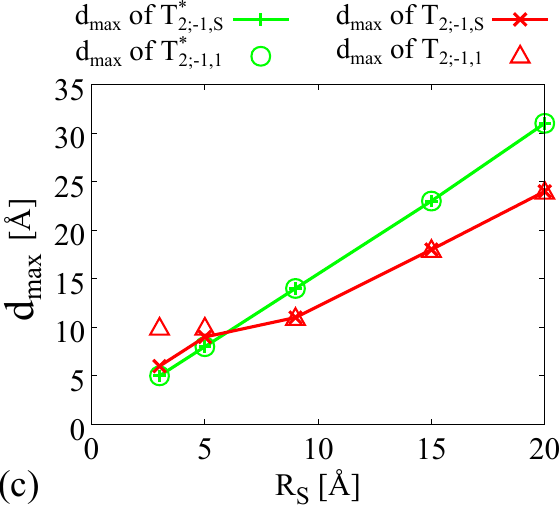}
\end{array}$\caption{Dependence of $T_2$ and $T_2^*$ on the distance $d$ between the two electron spins are shown in (a) for $R_S=5\AA$ and in (b) for $R_S=9\AA$, with other model parameters fixed. 
The value of $d$ where the coherence time reaches a maximum, $d_{max}$, as a function of $R_S$ is plotted in (c).
Other parameters of the model used for results in this figure are $J=10\GHz$,  $R_B=2\ang$, $n_B=0.01/\ang^{3}$, $B=1\T$, $\vec{d}\parallel\vec{B}$.
}\label{fig:T2Star_T2_vs_dz}
\end{figure}

The effect of tuning the distance between the two electron spins $d$ is shown in Figs.~\ref{fig:T2Star_T2_vs_dz}(a) and (b) for two different $R_S$ while other parameters are fixed. 
Both $T_2^*$ and $T_2$ show an optimum regime at $d$ values {\color{black} comparable to $2R_S$} and approach a constant at large $d$. 
The positions of the peaks $d_{max}$, defined as the value of $d$ where the coherence time reaches a maximum, are {\color{black} comparable to} and depend on $R_S$, as shown in Fig.~\ref{fig:T2Star_T2_vs_dz}(c). 
$d_{max}$ for each of $T^*_{2;-1,S}$, $T^*_{2;-1,1}$, $T_{2;-1,S}$ and $T_{2;-1,1}$ increase with $R_S$. 
{\color{black}
The physical understanding of the $d$ dependence will be detailed in Sect.~\ref{subsec:understanding}.
}

\begin{figure}[htp]
$\begin{array}{cc}
\includegraphics[width=0.48\columnwidth]{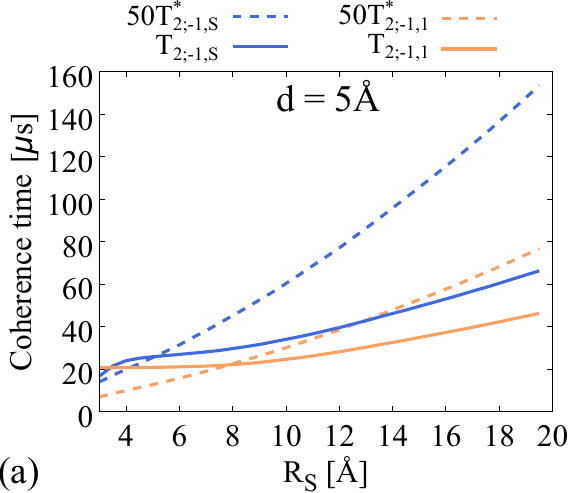}
~~~~~
\includegraphics[width=0.48\columnwidth]{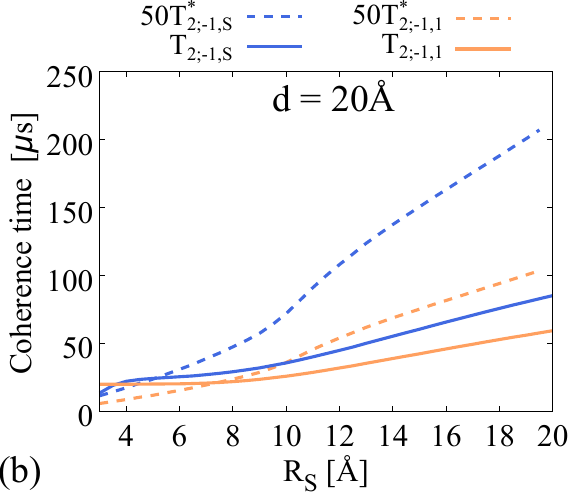}
\end{array}$\caption{Dependence of $T_2$ and $T_2^*$ on the minimum distance between an electron and the closest proton, $R_S$, for (a) $d=5\AA$ and (b) $d=20\AA$.  Other parameters of the model used for results in this figure are $J=10\GHz$, $R_B=2\ang$, $n_B=0.01/\ang^{3}$, $B=1\T$, $\vec{d}\parallel\vec{B}$.
}\label{fig:T2Star_T2_vs_r}
\end{figure}

In Figs.~\ref{fig:T2Star_T2_vs_r}(a) and (b), we show the dependence of $T_2$ and $T_2^*$ on $R_S$ for two different $d$ with other parameters fixed. All coherence times increase with $R_S$ as a result of a decrease in the maximum dipolar hyperfine interaction between electron and proton spins {\color{black} and thus the maximum of the fields $\left|\nabla_{\vec{r}}\left(A_{1nzz}+A_{2nzz}\right)\right|$ and $\left|A_{1nzz}+A_{2nzz}\right|$, which constrains decoherence due to any proton-pair flip-flop and inhomogeneous dephasing due to any single proton, respectively, according to Sects.~\ref{subsec:PCA_theory} and \ref{subsec:FID_theory}}. Within the range of $R_S$ of interest, $T_2^*$ is always sensitive to change in $R_S$. In the case of Hahn echo, $T_{2;-1,1}$ remains relatively constant at small $R_S$ until $R_S$ increases over a threshold independent of $d$ where $T_{2;-1,1}$ starts to increase noticeably. This is similar to the nuclear spin diffusion barrier in the context of single-electron spin decoherence\cite{graham2017synthetic,chen2020decoherence} and implies a barrier independent of $d$ around each electron spin. 
{\color{black}
This is reasonable as the decoherence $L^1_{-1,1}$  is similar to the Hahn-echo decoherence for single electron spins in the sense that the $\pi_{xx}$ pulse in our model flips the states $\left|E_{-1}\right\rangle$ and $\left|E_{1}\right\rangle$.
The existence of the barrier is correlated with the fact that a proton-spin pair that is too close to an electron spin has the gradient field $\left|\nabla_{\vec{r}}\left(A_{1nzz}+A_{2nzz}\right)\right|$ at its location and the $\left|E_{-1,k}\right|$ being too large so that its pair contribution to decoherence $g_k$ becomes negligibly small again, as shown in Fig.~\ref{fig:fk_gk_vs_C_E}(b).
}

\begin{figure}[htp]
$\begin{array}{c}
\includegraphics[width=0.5\columnwidth]{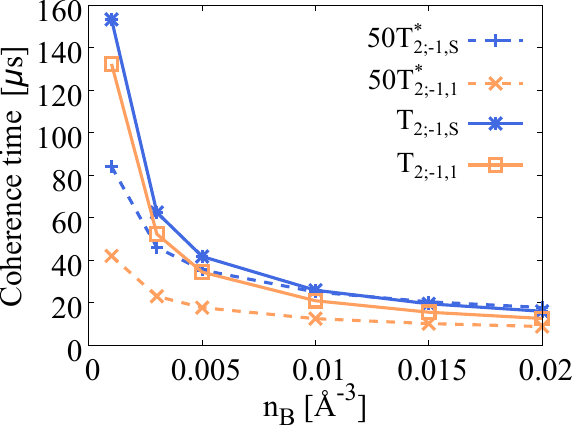}
\end{array}$\caption{Dependence of $T_2$ and $T_2^*$ on the proton density $n_B$. Other parameters of the model used for results in this figure are $J=10\GHz$, $d=5\ang$, $R_S=5\ang$, $R_B=2\ang$, $B=1\T$, $\vec{B}\parallel\vec{d}$.
}\label{fig:T2Star_T2_vs_np}
\end{figure}

When the density of protons $n_B$ is increased with other parameters fixed, $T_2$ and $T_2^*$ decrease, as shown in Fig.~\ref{fig:T2Star_T2_vs_np}. This is expected as the presence of more nearby protons produces a larger magnetic noise field on the electron spins. 
{\color{black} Specifically, for the scenario of $N=1$, increasing $n_B$ leads to more proton pair flip-flops $k$ in the products in Eqs.~\ref{eq:L1_-1_0_PCA_k_resolution},\ref{eq:L1_-1_1_PCA_k_resolution},\ref{eq:L1_-1_S_PCA_k_resolution},\ref{eq:L1_S_1_PCA_k_resolution} and \ref{eq:L1_0_1_PCA_k_resolution}; for the scenario of $N=0$, more protons lead to a larger inhomogeneous broadening, expressed in Eq.~\ref{eq:sigma_variance}, in the Overhauser field on electron spin system from pure states of protons.
}

\begin{figure}[htp]
$\begin{array}{c}
\includegraphics[width=0.6\linewidth]{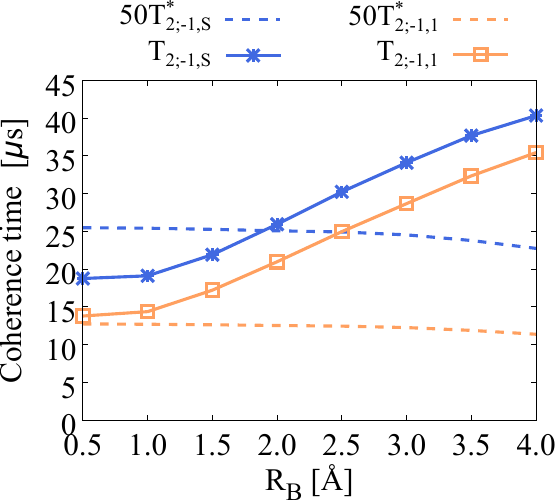}
\end{array}$
\caption{
Dependence of $T_2$ and $T_2^*$ on the minimum distance between protons, $R_B$.
Other parameters of the model used for results in this figure are $d=5\ang$, $R_S=5\ang$, $n_B=0.01/\ang^{3}$, $J=10\GHz$, $B=1\T$, $\vec{B}\parallel\vec{d}$.} \label{fig:T2Star_T2_vs_s}
\end{figure}

In Fig.~\ref{fig:T2Star_T2_vs_s}, we show the dependence of coherence times on the minimum spacing $R_B$ between proton spins.
In the {\color{black} scenario of FID}, the corresponding coherence times $T_2^*$ are largely insensitive to $R_B$. This is because the inhomogeneous dephasing $T_2^*$ in our model is the result of averaging the electron spin coherence over all possible proton initial pure states and not a result of the proton-proton interaction. The latter is also manifested in that for the simulation of FID, gCCE converges at order 1. {\color{black} The inhomogeneous broadening in the Overhauser field, Eq.~\ref{eq:sigma_variance} is insensitive to $R_B$ unless $R_B$ is so large that it starts to deplete the number of protons that are in close proximity to the electron spins, even though the overall average density $n_B$ across the entire space remains unchanged.
The slight decrease in $T_2^*$ at large $R_B$ in Fig.~\ref{fig:T2Star_T2_vs_s} is a consequence of this depletion.
} 
With the application of the Hahn-echo pulse, the  {\color{black} coherence times} of the two coupled electron spins now increase with the minimum spacing between proton spins, which sets the upper bound on the strength of the dipolar interaction {\color{black}$\left|C_{\alpha k}\right|=\left|d_{nm}\right|$} between nuclear spins and consequently the rate of nuclear spin flip-flop processes.
Notice that if two nuclear spins reside very close to each other, although the rate of the flip-flop process is high, the magnetic dipolar fields from them on each electron spin tend to cancel out, {\color{black} manifested as a vanishing $\left|E_{-1,k}\right|$ that results in a zero contribution to decoherence $f_k$ and $g_k$ as analyzed in Sect.~\ref{subsec:PCA_theory}}.  This leads to the relative insensitivity of $T_2$ to very small $R_B$ seen in Fig.~\ref{fig:T2Star_T2_vs_s}.

\begin{figure}[htp]
$\begin{array}{c}
\includegraphics[width=0.6\columnwidth]{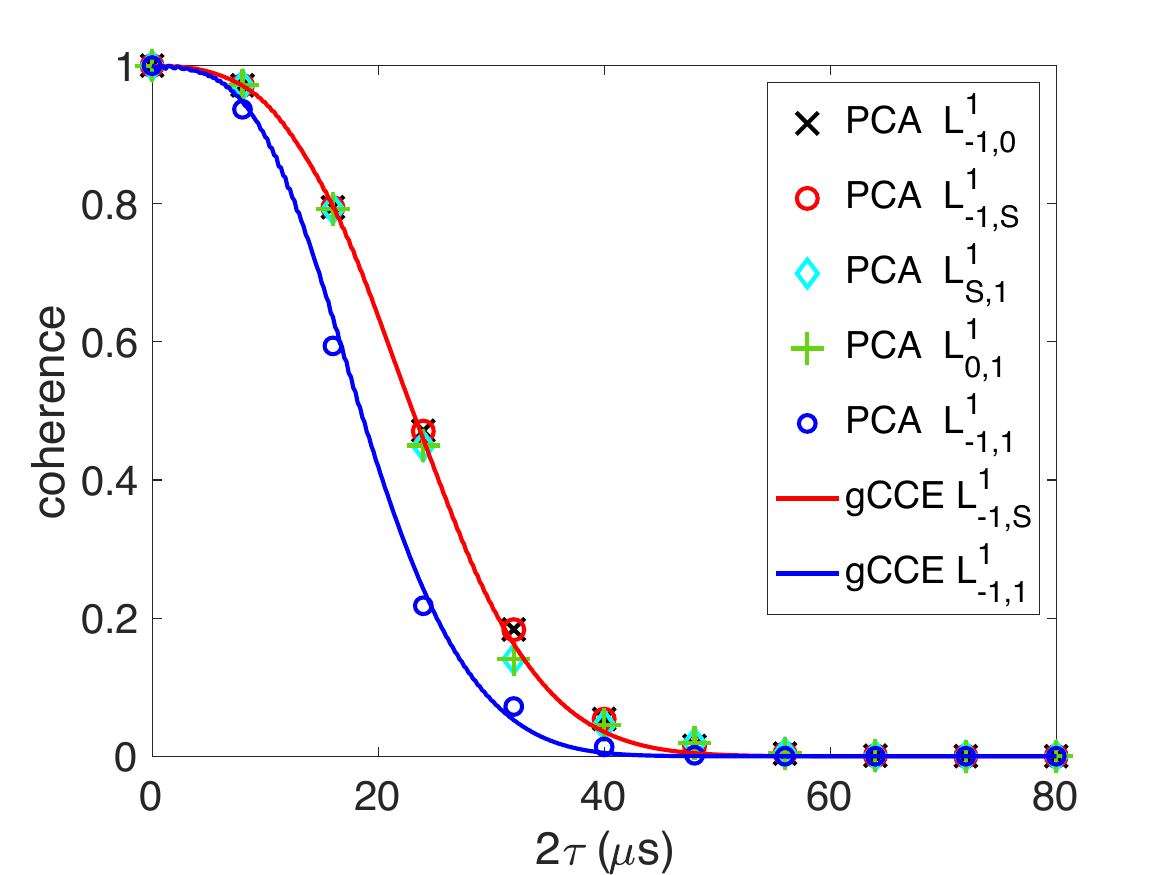}
\end{array}$\caption{{\color{black} The coherence functions $L^{1}_{\alpha\beta}$ calculated with Eqs.~\ref{eq:L1_-1_0_PCA_k_resolution}--\ref{eq:L1_0_1_PCA_k_resolution} within PCA for the regime of large $B$ and $J$ agree well with the gCCE results. Note that both PCA and gCCE $L^1_{0,S}$ equal one on the time scale of the plot and are not shown. The coherence functions in this plot are calculated for a large bath with a single random spatial configuration, with parameters $B=1\T$, $\vec{d}\parallel\vec{B}$, $J=10\GHz$, $d=10\ang$, $R_S=5\ang$, $R_B=2\ang$, $n_B=0.01/\ang^{3}$.
}}\label{fig:PCA_gCCE_agree}
\end{figure}

{\color{black} 
\subsection{Understanding dependence of coherence times on field orientation and $d$ }\label{subsec:understanding}
In order to understand the cause of the field orientation dependence of coherence times, for the scenario of $N=1$, we adopt the theory of PCA applied to our model as presented in Sect.~\ref{subsec:PCA_theory}. 
The PCA is valid for our model in the regime of large field and exchange and agrees with the gCCE result, as shown in Fig.~\ref{fig:PCA_gCCE_agree}.
In Sect.~\ref{subsec:PCA_theory} PCA predicts that nuclear spin pairs that contribute the most to decoherence are those with properties (1) the position vector $\vec{R}_{nm}$ joining the two protons is short and is either nearly parallel or nearly perpendicular to the external field, and (2) Viewing secular hyperfine interaction as a function of the proton position $\vec{r}$, its gradient field $\nabla_{\vec{r}}( A_{1zz}+A_{2zz})|_{\vec{r}=\vec{R_c}}$  at the location (of the center $\vec{R_c}$) of the pair is large enough and nearly parallel to $\vec{R}_{nm}$. 
Property (1) is confirmed by studying the length of $\vec{R}_{nm}$ and the angle between $\vec{R}_{nm}$ and $\vec{B}$ of the nuclear spin pairs that contributes the most to the decoherence. Fig.~\ref{fig:pair_angle_histogram} shows some statistics of the nuclear spin pairs in a single random configuration that contribute decoherence $f_k$ to $L^1_{-1,S}$, as defined in Sect.~\ref{subsec:PCA_theory}, that is greater than $0.001$ at a time $2\tau$ comparable to $T_{2;-1,S}$. Figs.~\ref{fig:pair_angle_histogram}(a) and (b) show the projection onto the $xz$ plane, for viewing purposes, of $\vec{R}_{nm}$'s that join pairs of protons as red line segments, for the situations of $\vec{B}\parallel\vec{d}$ and $\vec{B}\perp\vec{d}$, respectively. The two black dots mark the locations of the electron spins, which are on the $xz$ plane. The positions of all the spins in Fig.~\ref{fig:pair_angle_histogram}(b) are rotated from those in Fig.~\ref{fig:pair_angle_histogram}(a) about the $y$ axis by $90\degree$, with the field direction always along $z$. 
These $\vec{R}_{nm}$ are short in the sense that they mainly connect to nearest neighbors. The distribution of the angle $\theta_{nm}$ between $\vec{R}_{nm}$ and $\vec{B}$ of the pairs in Fig.~\ref{fig:pair_angle_histogram}(a) and (b) are shown in Fig.~\ref{fig:pair_angle_histogram}(c) and (d), respectively. It is obvious that the distribution consists of two major parts. One part is a peak around $90\degree$ that represents pairs with $\vec{R}_{nm}$ nearly perpendicular to the field. The other part is centered around $0\degree$/$180\degree$ representing pairs with $\vec{R}_{nm}$ nearly parallel to the field. The minimum of distribution near $54.7\degree$/$125.3\degree$ corresponds to the pair orientation where the dipolar interaction $d_{nm}$ vanishes (Eq.~\ref{eq:d_nm}). 
The dip at $0\degree$/$180\degree$ in the distribution is the result of a much smaller solid angle corresponding to the same $\Delta\theta$ window at $0\degree$/$180\degree$ than at other angles. Similar results as in Fig.~\ref{fig:pair_angle_histogram} can also be obtained for the pair contribution of decoherence $g_k$ to $L^1_{-1,1}$.

\begin{figure}[htp]
$\begin{array}{c}
\includegraphics[width=0.9\columnwidth]{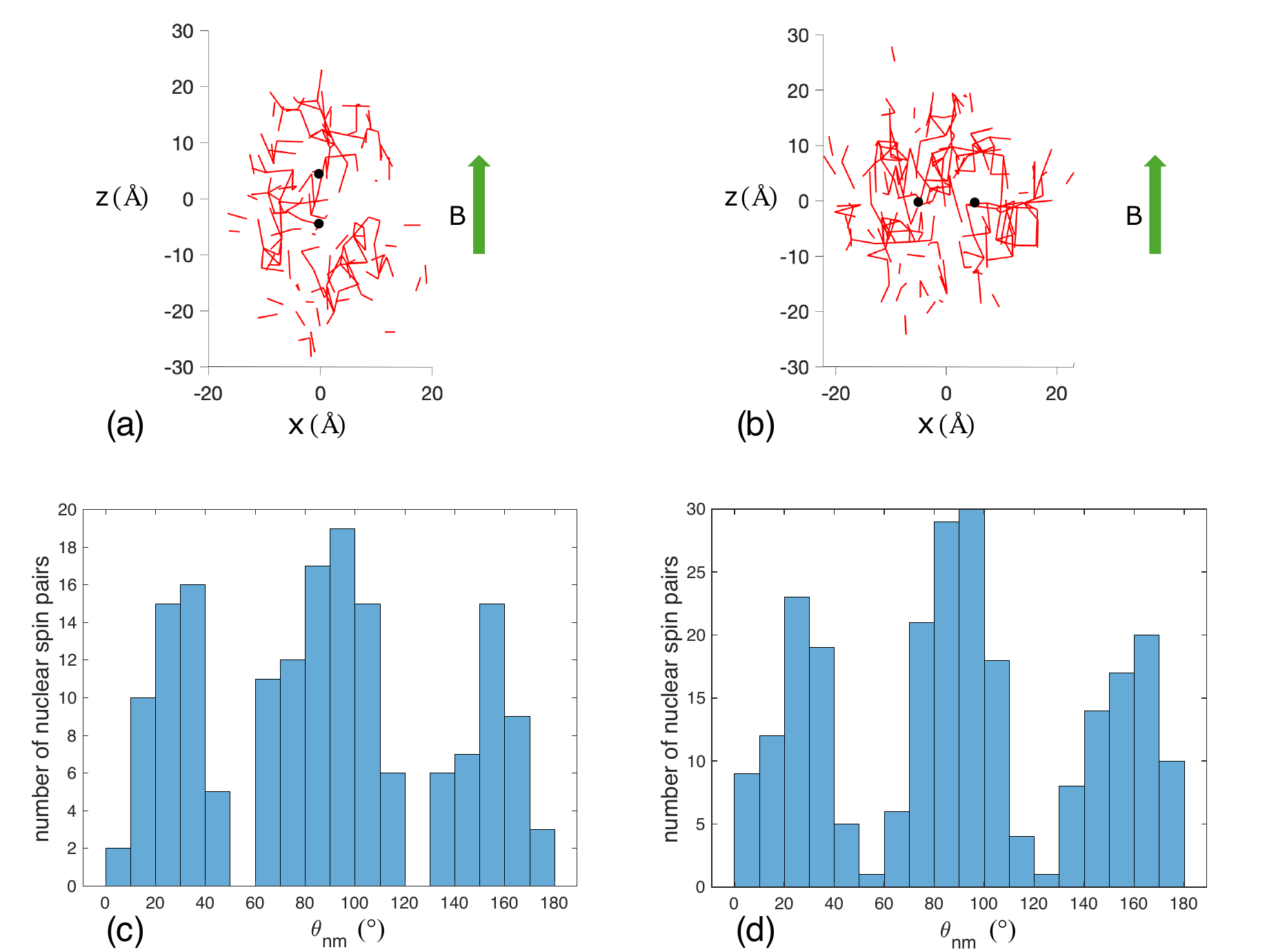}
\end{array}$\caption{{\color{black} Projection onto $xz$ plane of $\vec{R}_{nm}$ of the proton pairs that contributes decoherence $f_k>0.001$ for a single proton spatial configuration when (a) $\vec{B}\parallel\vec{d}$ and (b) $\vec{B}\perp\vec{d}$.  
The two black dots mark the locations of the electron spins on the $xz$ plane. The positions of the all spins in (b) are rotated from (a) about the $y$ axis by $90\degree$, with the field direction always along $z$.
(c) and (d) show the distribution of the angle $\theta_{nm}$ between $\vec{R}_{nm}$ and $\vec{B}$ of the pairs in Fig.~\ref{fig:pair_angle_histogram}(a) and (b), respectively.
Model parameters used for this example are $B=1\T$, $J=10\GHz$, $d=10\ang$, $R_S=5\ang$, $R_B=2\ang$, $n_B=0.01/\ang^{3}$, $2\tau=40\mus$.
}}\label{fig:pair_angle_histogram}
\end{figure}

The total number of pairs, which contributes $f_k>0.001$, is $168$ in Fig.~\ref{fig:pair_angle_histogram}(a) and (c) and $247$ in Fig.~\ref{fig:pair_angle_histogram}(b) and (d). This larger number of pair of protons that contribute to decoherence leads to a smaller $T_2$ when $\vec{B}\perp\vec{d}$ and originates from a larger spatial region where property (2) can be satisfied.
The gradient field $\nabla_{\vec{r}}( A_{1zz}+A_{2zz})$ on the $xz$ plane is plotted in Fig.~\ref{fig:gradient_A1nzz_A2nzz_B_orientation}(a) and (b) for the positions of electron spins and the field direction in Fig.~\ref{fig:pair_angle_histogram}(a) and (b), respectively. The empty region around each electron spin (black dot) denotes the radius $R_S$ free of protons. Note that this gradient field due to two electron spins is a superposition of the field due to each of them $\nabla_{\vec{r}} A_{izz}$ $(i=1,2)$, which has a cylindrical symmetry about the axis through the spin in the field direction $z$.
The regions where $\nabla_{\vec{r}}( A_{1zz}+A_{2zz})$ is the largest in magnitude and is parallel or perpendicular to $\vec{B}$, as a requirement of property (2) that the gradient field be parallel to $\vec{R}_{nm}$, are marked by red circles.  
Note that the large field vectors at $z=0$ in Fig.~\ref{fig:gradient_A1nzz_A2nzz_B_orientation}(a) are not marked as they reside in the small volume between the two spheres free of protons, so the number of pairs populated there is small.
As shown in Fig.~\ref{fig:gradient_A1nzz_A2nzz_B_orientation}, two such regions in (a) while four in (b), indicating that the pairs that are populated in such regions and cause the most electron spin decoherence are more when $\vec{B}\perp\vec{d}$.
Obviously, this field orientation dependence of $T_2$ is the most prominent when $d$ is comparable to $2R_S$ where the region between the electron spins is forbidden to protons, and it is a consequence of anisotropy of the gradient field $\nabla_{\vec{r}} A_{izz}$ $(i=1,2)$ around a single electron spin. 
The same argument here applies to the field orientation dependence of $L^1_{-1,1}$ as well.

\begin{figure}[htp]
$\begin{array}{c}
\includegraphics[width=0.7\columnwidth]{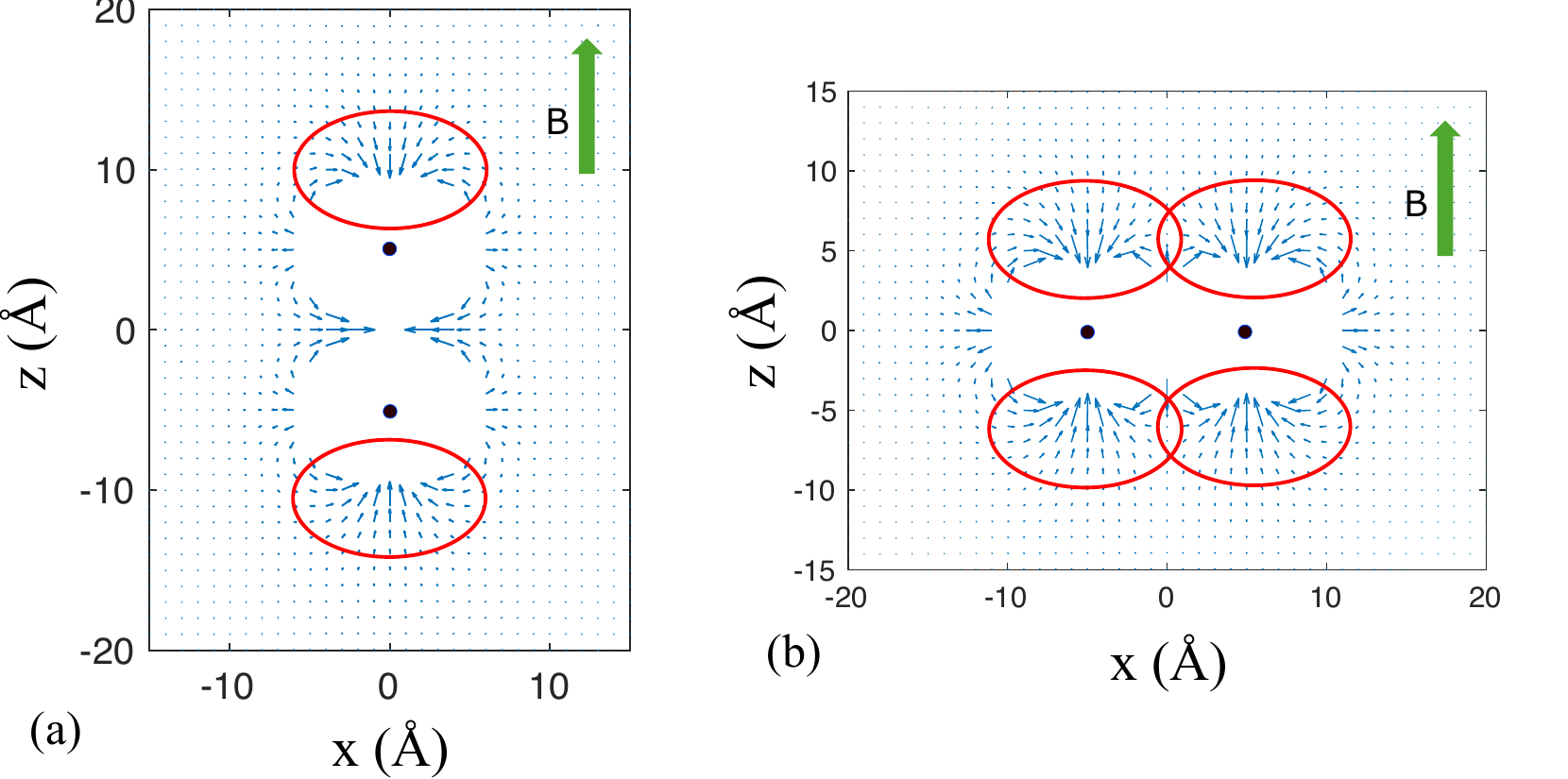}
\end{array}$\caption{{\color{black} $\nabla\left(A_{1nzz}+A_{2nzz}\right)$ in the $xz$ plane plotted for different field orientations relative to $\vec{d}$, (a) $\vec{B}\parallel \vec{d}$, (b) $\vec{B}\perp \vec{d}$. Blue arrows in both plots are the gradient field and share the same unit length.
The empty area near the center where no field vector is given represents the region free of protons.
Black dots mark positions of the electron spins in the $xz$ plane. Red circles mark the regions in which $\nabla\left(A_{1nzz}+A_{2nzz}\right)$ satisfy the property (2) of the nuclear spin pairs contributing the most to decoherence.
Relevant parameters of the model used in this figure are $d=10\ang$, $R_S=5\ang$.
}}\label{fig:gradient_A1nzz_A2nzz_B_orientation}
\end{figure}

\begin{figure}[htp]
$\begin{array}{c}
\includegraphics[width=0.6\columnwidth]{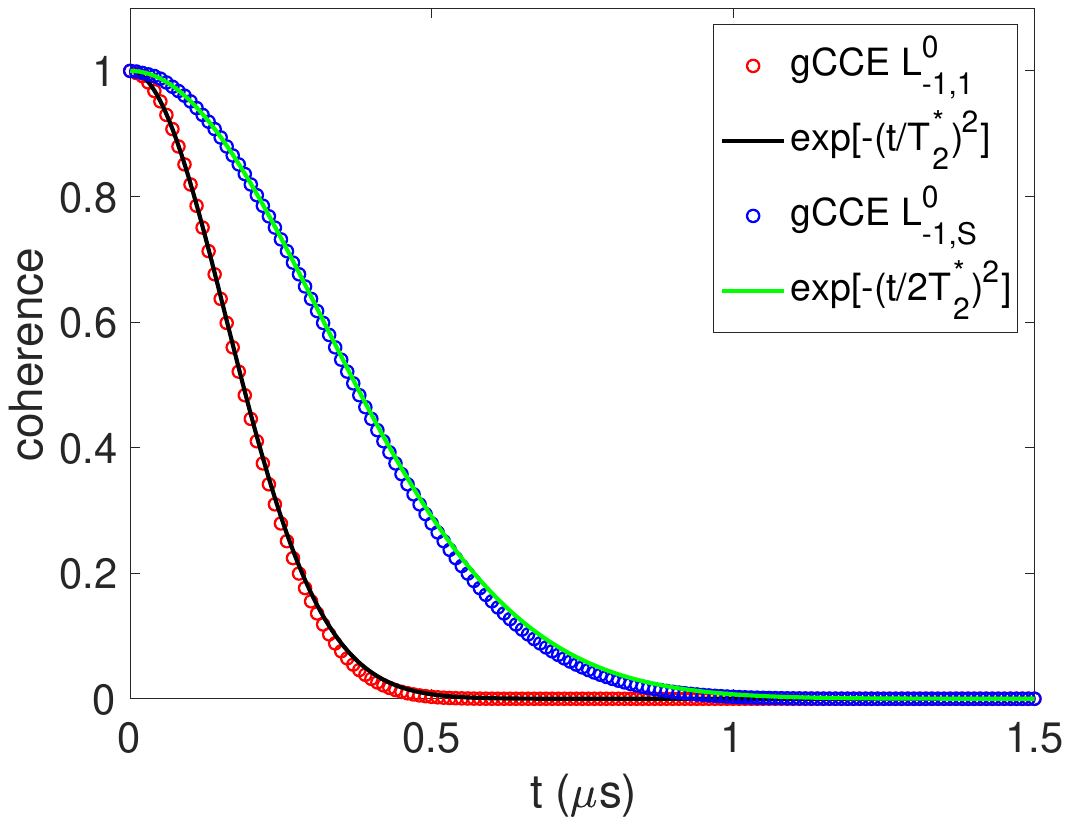}
\end{array}$\caption{{\color{black} The coherence functions $L^{0}_{\alpha\beta}$ calculated with Eqs.~\ref{eq:L0_-1_1} and \ref{eq:sigma_variance}--\ref{eq:T2_star_theory} for the regime of large $B$ and $J$ agree well with the gCCE results. Note that both theoretical and gCCE $L^0_{0,S}$ equal one on the time scale of the plot and are not shown. The coherence functions in this plot are calculated for a single random large bath, with parameters $B=1\T$, $\vec{d}\parallel\vec{B}$, $J=10\GHz$, $d=10\ang$, $R_S=5\ang$, $R_B=2\ang$, $n_B=0.01/\ang^{3}$.
}}\label{fig:FID_theory_gCCE_agree}
\end{figure}

The field orientation dependence of the $T_2^*$'s can be understood in a similar fashion, although one now needs to consider instead the scalar field  $\left|A_{1nzz}+A_{2nzz}\right|$,  according to the expression of $T_2^*$ (Eqs.~\ref{eq:sigma_variance}--\ref{eq:T2_star_theory}) in the regime of high field and exchange interaction. 
The agreement between the calculated gCCE $L^{0}_{-1,1}$ and $L^{0}_{-1,S}$ and the expressions Eqs.~\ref{eq:L0_-1_1} and \ref{eq:sigma_variance}--\ref{eq:T2_star_theory} is shown in Fig.~\ref{fig:FID_theory_gCCE_agree}.
$\left|A_{1nzz}+A_{2nzz}\right|$ has the same symmetry as its gradient considered above and essentially the same regions in space where its value is large. Therefore, the argument above on the different number of regions where the field magnitude is the largest can be applied here as well to explain that more protons (not pairs here) can be populated in such regions when $\vec{B}\perp\vec{d}$, compared to $\vec{B}\parallel\vec{d}$, leading to larger inhomogeneous broadening in the Overhauser fields and smaller $T_2^*$.
Fig.~\ref{fig:A1nzz+A2nzz_B_orientation} plots the $\left|A_{1nzz}+A_{2nzz}\right|$ in the $xz$ plane for both cases of $\vec{B}\parallel\vec{d}$ and $\vec{B}\perp\vec{d}$, of which the corresponding gradient $\nabla\left(A_{1nzz}+A_{2nzz}\right)$ has been shown in Fig.~\ref{fig:gradient_A1nzz_A2nzz_B_orientation}.

\begin{figure}[htp]
$\begin{array}{c}
\includegraphics[width=0.9\columnwidth]{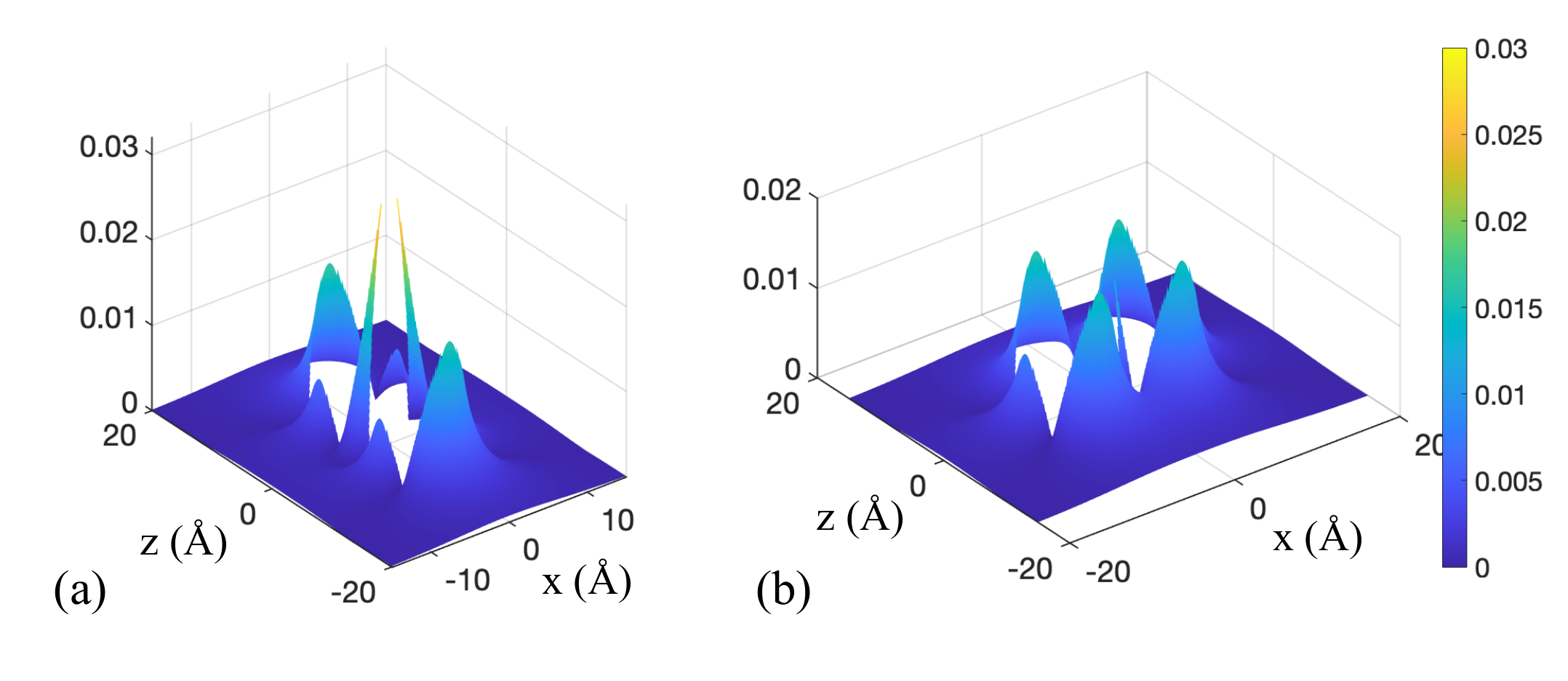}
\end{array}$\caption{{\color{black} $\left|A_{1nzz}+A_{2nzz}\right|$ in the $xz$ plane in the unit of $(-\gamma_{e}\gamma_{p}\mu_{0}\hbar/2)$ is plotted for different field orientations relative to $\vec{d}$, (a) $B\parallel d$, (b) $B\perp d$. The empty area near the center where no field value is given represents the region free of protons.
Relevant parameters of the model used in this figure are $d=10\ang$, $R_S=5\ang$.
}}\label{fig:A1nzz+A2nzz_B_orientation}
\end{figure}

In the following, we present the understanding of the dependence on $d$ of the coherence times as shown in Fig.~\ref{fig:T2Star_T2_vs_dz}.}
The coherence times are insensitive to change in large $d$ results because only protons within a finite radius $r_{\bath}$ 
of either electron spin can contribute to decoherence, as explained in Supplementary Material Sect.~S2. 
{\color{black} In terms of the fields $\left|A_{1nzz}+A_{2nzz}\right|$ and $\nabla_{\vec{r}}\left(A_{1nzz}+A_{2nzz}\right)$, as they both decay rapidly with the increasing distance $\left|\vec{r}\right|$ of the proton from the middle point between the electron spins (the former decays as $\left|\vec{r}\right|^{-3}$ and the later $\left|\vec{r}\right|^{-4}$), they both effectively vanish outside the radius $r_{bath}$ which renders that no proton spins outside can contribute to dephasing in $L^0$ or $L^1$.}
When the distance between the two electron spins is increased to more than twice of this radius, the total Hamiltonian effectively no longer changes as far as decoherence of the electron spins is concerned.
{\color{black} 
The existence of the optimum distance can be understood by again inspecting $\nabla_{\vec{r}}\left(A_{1nzz}+A_{2nzz}\right)$ and $\left|A_{1nzz}+A_{2nzz}\right|$, similar to the analysis of the field orientation dependence of coherence times. Taking the former as an example, $\nabla_{\vec{r}}\left(A_{1nzz}+A_{2nzz}\right)$ in the $xz$ plane is plotted in Fig.~\ref{fig:gradient_A1nzz_A2nzz_d_dependence} for the cases representing $d\gg2R_S$, $d\approx2R_S$, and $d\ll2R_S$. As shown in Fig.~\ref{fig:gradient_A1nzz_A2nzz_d_dependence}(a), in four regions in space enclosed by the red circles, $\nabla_{\vec{r}}\left(A_{1nzz}+A_{2nzz}\right)$ can reach the largest magnitude and satisfy property (2) for nuclear spin pairs causing the most decoherence. As $d$ decreases to be comparable to $2R_S$ such that the space between the electron spins where protons can be populated is eliminated, only two such regions remain, the same as the condition in Fig.~\ref{fig:gradient_A1nzz_A2nzz_B_orientation}(a). Therefore, fewer proton spin pairs cause decoherence, and the coherence times $T_{2;-1,S}$ and $T_{2;-1,1}$ are greater than in the large $d$ regime. As $d$ becomes much smaller than $2R_S$, the local gradient fields $\nabla_{\vec{r}}A_{inzz}$ around each electron spin superpose constructively, causing an increasing magnitude of $\nabla_{\vec{r}}\left(A_{1nzz}+A_{2nzz}\right)$ around the electrons, especially in the two circled regions. As a result, proton spin pairs cause stronger decoherence compared to $d\approx2R_S$.
Similar argument can be made about $\left|A_{1nzz}+A_{2nzz}\right|$ to understand $d_{max}$ for $T^*_{2;-1,S}$ and $T^*_{2;-1,1}$, which is obvious from the plots of the scalar field for $d\gg2R_S$, $d\approx2R_S$ and $d\ll2R_S$ shown in Fig.~\ref{fig:A1nzz_A2nzz_d_dependence}.
In summary, the optimal distance is sufficiently short, so that the middle region between the two electron spins is forbidden to the protons but also sufficiently long, so that each of the nearby protons strongly interacts with only one of the electrons.

}

\begin{figure}[htp]
$\begin{array}{c}
\includegraphics[width=0.7\columnwidth]{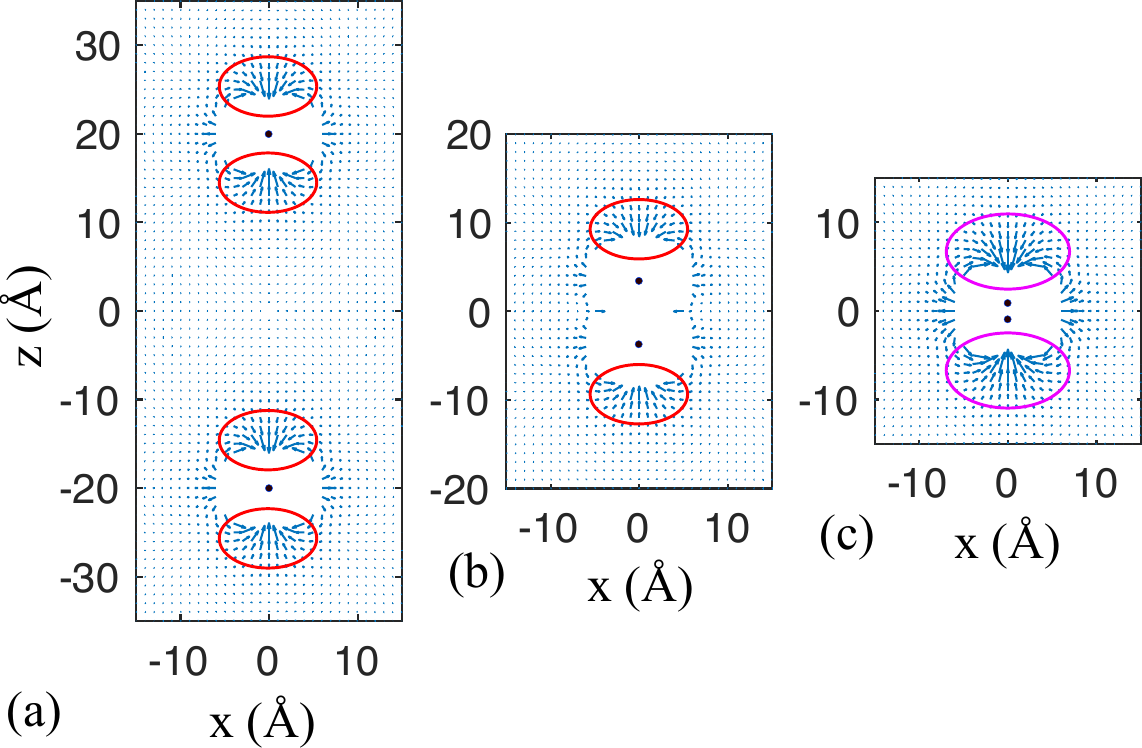}
\end{array}$\caption{{\color{black} $\nabla_{\vec{r}}\left(A_{1nzz}+A_{2nzz}\right)$ in the $xz$ plane is  plotted for (a) $d=40\ang$, (b) $d=7\ang$ and (c) $d=2\ang$, representing the regimes of $d\gg2R_S$, $d\approx2R_S$ and $d\ll2R_S$, respectively.
The gradient field vectors in these subplots share the same unit length. 
Black dots mark locations of electron spins.
Red and purple circles mark the regions where the gradient field has large strength and satisfy property (2) of nuclear spin pairs contributing the most to decoherence.  
Parameters of the model used in this figure are $R_S=5\ang$,  $\vec{d}\parallel\vec{B}$.
}}\label{fig:gradient_A1nzz_A2nzz_d_dependence}
\end{figure}

\begin{figure}[htp]
$\begin{array}{c}
\includegraphics[width=0.95\columnwidth]{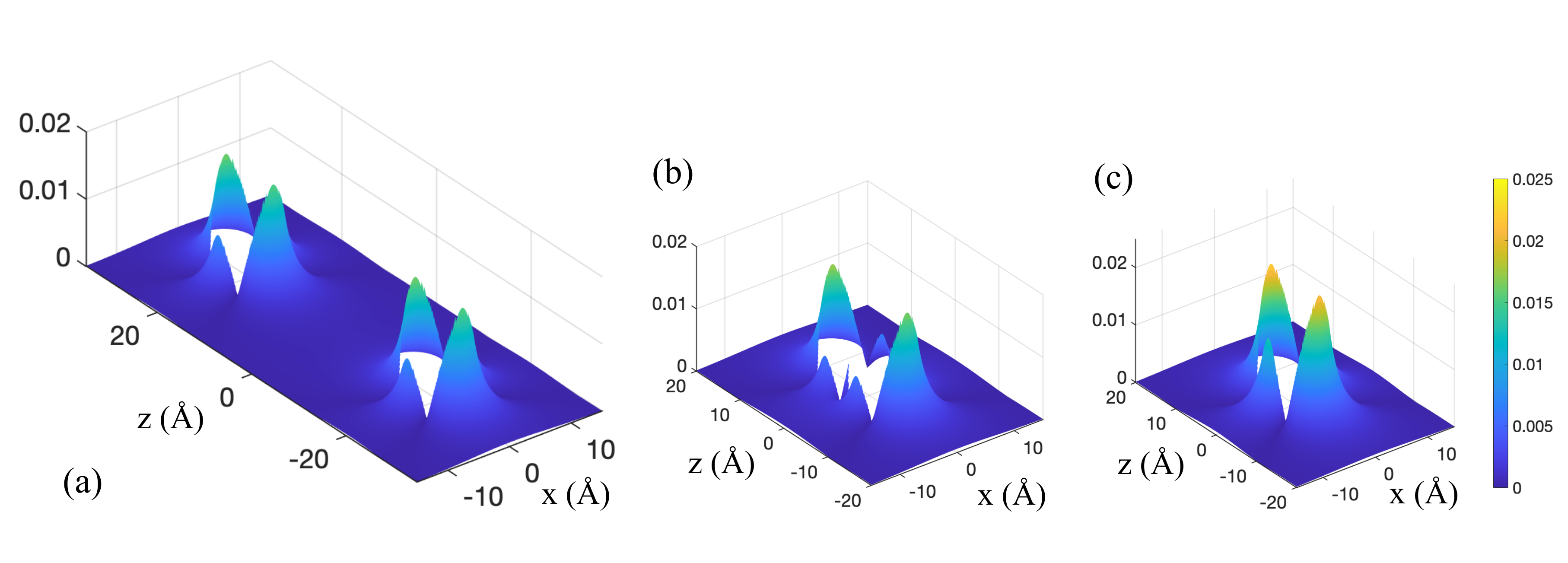}
\end{array}$\caption{{\color{black} $\left|A_{1nzz}+A_{2nzz}\right|$ in the unit of $(-\gamma_{e}\gamma_{p}\mu_{0}\hbar/2)$ in the $xz$ plane is plotted for (a) $d=40\ang$, (b) $d=7\ang$ and (c) $d=2\ang$, representing the regimes of $d\gg2R_S$, $d\approx2R_S$ and $d\ll2R_S$, respectively. 
Parameters of the model used in this figure are $R_S=5\ang$,  $\vec{d}\parallel\vec{B}$.
}}\label{fig:A1nzz_A2nzz_d_dependence}
\end{figure}

{\color{black}
\subsection{Implication for experiments}\label{subsec:implications}
}

{\color{black}The evolution of the system in the form of Eq.~\ref{eq:unitary_evolution} is relevant to several possible experiments. In the case of electron spins performing an entangling two-qubit gate utilizing the isotropic exchange coupling, the $N=0$ scenario corresponds to, for example, an $\sqrt{SWAP}$ gate on the computation basis chosen as $|\uparrow_j\rangle$ 
and $|\downarrow_j\rangle$ for each spin, given that $J\cdot t=(2n+1)/4$ and $\gamma_{e}B\cdot t=m/2$ with $m,n$ being any integer. Then the $N=1$ scenario, with an additional $\pi_{xx}$ pulse added at the end of the evolution, i.e. at time $t$, retains the $\sqrt{SWAP}$ behavior of the two-qubit gate, as long as $J\cdot t=(2n+1)/4$ is still satisfied, while mitigating noises that couple to $\hat{S}_{1z}$ and $\hat{S}_{2z}$\cite{viola1998dynamical} for a higher gate fidelity. This additional $\pi_{xx}$ does not change the value of the off-diagonal elements of the two-electron RDM in energy basis but only switches them. Another case which is more recently accessible is the pulse electron paramagnetic resonance (EPR) experiment, where decay of transverse magnetization is measured. 
$N=0$ and $N=1$ scenarios above simulate FID and Hahn echo experiments in standard pulse EPR, respectively.
The transverse magnetization is closely related to the off-diagonal elements of the two-electron RDM. For example, if the two electrons have the same isotropic $g$ factor and are coupled with an isotropic exchange,  and the first $\pi/2$ pulse in a pulse EPR experiment prepares the two electrons spins in the transverse plane, with the pure state $\frac{1}{2}(\left|\uparrow_1\rangle+|\downarrow_1\right\rangle)\otimes(\left|\uparrow_2\rangle+|\downarrow_2\right\rangle)$ for the initial RDM $\hat{\rho}_{S}(0)$, one can show that the transverse magnetization $\left|\Tr[\hat{\rho}_{S}(t)\hat{S}_{tot}^{-}]\right|=\sqrt{2}\left|\langle E_0\right|\hat{\rho}_{S}(t)\left|E_{-1}\rangle+\langle E_1\right|\hat{\rho}_{S}(t)\left|E_0\rangle\right|$
, where $\hat{S}_{tot}^{-}=(\hat{S}_{1x}+\hat{S}_{2x})-i(\hat{S}_{1y}+\hat{S}_{2y})$. 
A third possible case is the two-electron-spin encoding of a single qubit, where any two energy states in Eq.~\ref{eq:isotropic_exchange_eigenenergy_basis} can be used as the computational basis. 
$N=0$ scenario in our model corresponds to the free evolution of the qubit. 
$N=1$ scenario corresponds to an effective Hahn echo experiment performed on the two-level system of
$\left|E_{-1}\right>$ and $\left| E_{1} \right>$, 
as the $\pi_{xx}$ pulse flips these two levels while leaving the other two unchanged. 
Finally, we point out that one way the state of the two-spin system can in principle be charecterized experimentally is to first apply a large enough magnetic field gradient to detune the two spins and measure each of the spins separately\cite{thalineau2014interplay,otsuka2016single,kandel2021perspective}.
}

Now we discuss the implications from the results {\color{black} in Sect.~\ref{subsec:gCCE results} and \ref{subsec:understanding}} for the experimental realization of magnetic molecular two-qubit systems with longer coherence time. 
Firstly, the Hahn-echo dynamical decoupling pulse, which is the same as the one used in the experiments for single electron spins except that now it acts on each of the coupled two spins at the same time, can enhance the coherence time by an order of magnitude, given that the pulse is ideal. {\color{black} This is beneficial for realizing entangling two-qubit gates with longer coherence time and higher fidelity. An example is the $\sqrt{SWAP}$ gate as mentioned in Sect.~\ref{subsec:Decoherence}. } In experiments, the Hahn echo coherence time $T_2$ as well as the fidelity of two-qubit gates can be further decreased by imperfection of pulses, and dynamical decoupling is generally harder to realize than the bare free evolution in FID. Therefore, both FID and Hahn-echo scenarios are worth study experimentally for magnetic molecular two-qubit systems.

For two molecular spin-$\frac{1}{2}$ qubits coupled by an isotropic exchange interaction, if the distribution of nearby protons {\color{black} resembles a random configuration, in the sense that protons are  distributed roughly evenly in space and the orientation of the vector $\vec{R}_{nm}$ joining neighboring protons does not strongly favor any particular direction,}  except for a radius around each qubit that is free of protons, then aligning the external magnetic field with the line joining the two qubits results in a coherence time higher than the orthogonal orientation. 
{\color{black} If the nearby protons do not resemble a random configuration, then an optimum direction of the magnetic field can still be found. 
In the cases of strong anisotropy in the spatial distribution of nearby protons or in the orientation distribution of nearby proton pairs, stronger field-orientation-dependent decoherence can occur.
For Hahn-echo experiments, one can orient the magnetic field ($z$ axis) so that the gradient field of the total secular hyperfine interaction $\nabla_{\vec{r}}\left(A_{1nzz}+A_{2nzz}\right)$ is small at the locations where the protons are densely distributed or where few pairs of neighboring protons are near parallel or perpendicular to the magnetic field. The guiding principle is to reduce the number of proton pairs that satisfy the two properties mentioned in Sect.~\ref{subsec:PCA_theory}.
For FID experiments, one should orient the magnetic field such that the scalar field $\left|A_{1nzz}+A_{2nzz}\right|$ is small at locations of most protons.}

It is in general favorable for longer coherence to use a large enough external magnetic field, optimally to work in the regime where the relevant coherence time is no longer sensitive to change in the field strength. 
Large fields suppress the hyperfine-mediated interaction between nuclear spins (the terms in {\color{black} the second row of} Eq.~\ref{eq:QDPT_H}, where the denominator $E_\alpha-E_\beta$ can be proportional to the field strength), which is known to reduce $T_2$ of single electron spins at relatively small fields\cite{liu2007control,cywinski2009pure,neder2011semiclassical}, {\color{black} and suppress the nonsecular part of the hyperfine interaction and of the dipolar interaction between protons.}

Changing the strength of the exchange interaction within the many orders of magnitudes typical for magnetic molecules essentially does not affect the coherence times if other physical quantities in the spin Hamiltonian describing the electron-nuclear system remain the same. 
{\color{black}
Therefore, in designing the exchange interaction of two coupled molecular spin qubits, the goal should not be to maximize coherence times but rather to ensure that the timescale of the intended two-qubit operations is well within the available coherence window. In the example of the $\sqrt{SWAP}$ gate, this timescale is of the order of $J^{-1}$.
}

For the spatial proton distribution described above that {\color{black} resembles a random configuration} in the space outside of a radius $R_S$ free of protons around each qubit, there is an optimal distance $d$ between the two qubits of the same order as $R_S$ that results in the longest coherence time, if $R_S$, the proton density and the minimum spacing between the protons are not affected by the tuning of $d$. 
Therefore, it is also beneficial for longer coherence time to optimize the qubit-qubit distance{\color{black}, e.g. by changing the linker fragment between two spin centers in a molecule}. 
{\color{black} For general systems, it is suggested to decrease the distance so that the region between two qubits is free of protons while avoiding the situation that some nearby protons strongly interact with both electron spins at the same time.}
Note that although the exchange interaction $J$ changes with $d$ between different realistic molecules, as long as $J$ is still within the pure dephasing regime, the above discussion of the effect of $d$ holds due to insensitivity of coherence times to $J$. 

{\color{black} As suggested from the dependence of the coherence times on $R_S$,} increasing the minimum distance between the electron and proton spins improves coherence, given that {\color{black} the proton spin configuration in space at greater distances is not affected}.  {\color{black} It should be noted that} in the scenario of Hahn-echo experiments, the increase of $T_{2{\color{black};-1,1}}$ only becomes significant when {\color{black} the minimum electron-proton distance} increases above the nuclear spin diffusion barrier.
The bath property of proton density has a great impact on the coherence times and it is always favorable to have a smaller density. In a situation where the proton density can not be varied, it is possible to further enhance the Hahn-echo coherence time if the minimum spacing between protons can be increased.

{\color{black}
\subsection{Discussion on dipolar interaction between electron spins}\label{subsec:dipolar}

So far we have been focusing on the situation where the isotropic exchange interaction $J$ is dominant and thus the only interaction included between the two electron spins. Before we conclude, in this subsection we discuss the effect of the magnetic dipolar interaction when it is not negligibly small compared to $J$ and emphasize that it leads to a stark difference in the energy states of the two electron spins, the dephasing dynamics of the two spins under the Hahn-echo pulse, and the stability of the gCCE method for some off-diagonal elements of the RDM.  

The electron-spin Hamiltonian of our model (Eq.~\ref{eq:H_S}), with the inclusion of the dipolar interaction, now reads
\bea
\hat{H}_{S}&=&-\gamma_{e}B\hat{S}^z_{1}-\gamma_{e}B\hat{S}^z_{2}+J\hat{\vec{S}}_{1}\cdot\hat{\vec{S}}_{2}+\hat{H}_{D},\label{eq:H_S_wDipole}
\eea
where $\hat{H}_{D}$ is the magnetic dipolar interaction between the two electrons, assumed to take the form for point dipoles,
\beq
\hat{H}_{D}=\hat{\vec{S}}_{1}\cdot\overleftrightarrow{P}\cdot\hat{\vec{S}}_{2},\label{H_D}
\eeq
\beq
\overleftrightarrow{P}=-\gamma_{e}^2\frac{\mu_{0}\hbar}{2d^{5}}\left[3\vec{d}\otimes\vec{d}-d^2\mathbb{I}_{3\times3}\right]\equiv -D\left[3\hat{d}\otimes\hat{d}-\mathbb{I}_{3\times3}\right] . \,
\label{eq:ee_dipolar}
\eeq
Without loss of generality, we assume that the position vector joining the two spins lies in the $xz$ plane. In addition, to fully show the difference caused by $\hat{H}_{D}$, we consider $\vec{d}$ at a general angle $\phi$ to the field $\vec{B}$ in the $z$ direction, $\vec{d}=d\hat{d}=d\left(\cos\phi\hat{z}+\sin\phi\hat{x}\right)$. $D=\gamma_{e}^2\frac{\mu_{0}\hbar}{2d^{3}}$ is the strength of the dipolar interaction. We will focus on the condition that the Zeeman energy $\left|\gamma_{e}B\right|\gg D$, since this represents the usual experimental situation where as the electron-electron distance $d$ is large enough so that $D$ is not negligible compared to $J$, $D$ is much smaller than the electron Zeeman energy. We will also again only consider the pure dephasing regime, where the difference between the energy gaps of $\hat{H}_S$ and the nuclear Zeeman energy is much larger than the dipolar hyperfine interaction, $\gamma_{e}\gamma_{p}\mu_{0}\hbar/2|\vec{r}_{jn}|^{3}\ll|\gamma_p B-|E_\alpha-E_\beta||$.

Now we examine the eigenstates of $\hat{H}_{S}$, only the superpositions of which will experience dephasing. Letting $-\gamma_{e}B\hat{S}^z_{1}-\gamma_{e}B\hat{S}^z_{2}\equiv \hat{H}_Z$ and $J\hat{\vec{S}}_{1}\cdot\hat{\vec{S}}_{2}\equiv\hat{H}_{J}$, we have $\hat{H}_{S}=\hat{H}_Z+\hat{H}_D+\hat{H}_J$. It is not difficult to show that $\left[\hat{H}_Z+\hat{H}_D,\hat{H}_J\right]=0$ and that the eigenstates of $\hat{H}_Z+\hat{H}_D$ are, in general, nondegenerate, so $\hat{H}_S$ and $\hat{H}_Z+\hat{H}_D$ share the same eigenstates. Even with this reduction, there is no simple closed-form expression for these states. The eigenstates derived symbolically using the Mathematica\cite{Mathematica} take the general form of
\bea
\left|E_S\right\rangle &=&\frac{1}{\sqrt{2}} \, (\left|\uparrow_1\rangle|\downarrow_2\right\rangle-\left|\downarrow_1\rangle|\uparrow_2\right\rangle),\nonumber\\
\left|E_{\widetilde{-1}}\right\rangle &=&a_{\widetilde{-1}}\left|\uparrow_1\rangle|\uparrow_2\right\rangle+b_{\widetilde{-1}}\left|\uparrow_1\rangle|\downarrow_2\right\rangle+b_{\widetilde{-1}}\left|\downarrow_1\rangle|\uparrow_2\right\rangle+c_{\widetilde{-1}}\left|\downarrow_1\rangle|\downarrow_2\right\rangle,\nonumber\\
\left|E_{\widetilde{0}}\right\rangle &=&a_{\widetilde{0}}\left|\uparrow_1\rangle|\uparrow_2\right\rangle+b_{\widetilde{0}}\left|\uparrow_1\rangle|\downarrow_2\right\rangle+b_{\widetilde{0}}\left|\downarrow_1\rangle|\uparrow_2\right\rangle+c_{\widetilde{0}}\left|\downarrow_1\rangle|\downarrow_2\right\rangle,\nonumber\\
\left|E_{\widetilde{1}}\right\rangle&=&a_{\widetilde{1}}\left|\uparrow_1\rangle|\uparrow_2\right\rangle+b_{\widetilde{1}}\left|\uparrow_1\rangle|\downarrow_2\right\rangle+b_{\widetilde{1}}\left|\downarrow_1\rangle|\uparrow_2\right\rangle+c_{\widetilde{1}}\left|\downarrow_1\rangle|\downarrow_2\right\rangle,\label{eq:H_SwD_eigenenergy_basis}
\eea
where $\left|\uparrow_i\right\rangle$ and $\left|\uparrow_j\right\rangle$ are the positive (negative) eigenstate of $\hat S^z_j$. 
The coefficients $a,b,c$ are such that the states are normalized, and they all in general depend on both $B$ and $D$.
Since one can show that $\hat{H}_S$ commutes with the total spin operator, $\left[\hat{H}_S,\left(\hat{\vec{S}}_{1}+\hat{\vec{S}}_{2}\right)^2\right]=0$, and in addition, we assumed $\left|\gamma_{e}B\right|\gg D$, consequently $\left|c_{\widetilde{-1}}\right|\approx1$, $\left|a_{\widetilde{-1}}\right|\approx1$,
$\left|b_{\widetilde{0}}\right|\approx 1/\sqrt{2}$, all other $a,b,c$ coefficients have a modulus much smaller than $1$, and we still label the three triplet states by $-1,0,1$ with a tilde indicating that they are now only approximate eigenstates of $\hat{S}_{1z}+\hat{S}_{2z}$.
The triplet states $\left|E_{-1,0,1}\right\rangle$ in Eq.~\ref{eq:isotropic_exchange_eigenenergy_basis} generally are no longer eigenstates of $\hat{H}_S$ when the dipolar interaction is present and are superpositions of $\left|E_{\widetilde{-1},\widetilde{0},\widetilde{1}}\right\rangle$ that now experience decay when pure dephasing between the latter occurs.
It is worth noting that when $\phi=0$, i.e. $\vec{d}\parallel\hat{z}$, the eigenstates after the inclusion of $\hat{H}_D$ remain those in Eq.~\ref{eq:isotropic_exchange_eigenenergy_basis}.
Simple close-form expressions also exist when $\phi=90\degree$, i.e. $\vec{d}\parallel\hat{x}$ and $\perp\vec{B}$, where 
\bea
\left|E_S\right\rangle &=&\frac{1}{\sqrt{2}} \, (\left|\uparrow_1\rangle|\downarrow_2\right\rangle-\left|\downarrow_1\rangle|\uparrow_2\right\rangle),\nonumber\\
\left|E_{\widetilde{-1}}\right\rangle &=&c_{\widetilde{-1}}\left(\frac{-4\left|B\gamma_e\right|+\sqrt{16B^2\gamma_e^2+9D^2}}{3D}\left|\uparrow_1\rangle|\uparrow_2\right\rangle+\left|\downarrow_1\rangle|\downarrow_2\right\rangle\right),\nonumber\\
\left|E_{\widetilde{0}}\right\rangle &=&\frac{1}{\sqrt{2}}\left(\left|\uparrow_1\rangle|\downarrow_2\right\rangle+\left|\downarrow_1\rangle|\uparrow_2\right\rangle\right),\nonumber\\
\left|E_{\widetilde{1}}\right\rangle&=&c_{\widetilde{1}}\left(-\frac{4\left|B\gamma_e\right|+\sqrt{16B^2\gamma_e^2+9D^2}}{3D}\left|\uparrow_1\rangle|\uparrow_2\right\rangle+\left|\downarrow_1\rangle|\downarrow_2\right\rangle\right).\label{eq:H_SwD_d_in_x_eigenenergy_basis}
\eea

The change in eigenstates due to dipolar interaction leads to different dephasing dynamics, especially in the Hahn-echo experiment. This can be best seen by considering the dynamics of the full electron-nuclei system under a projected Hamiltonian of the form of Eq.~(\ref{eq:pure_dephasing_total_H}), $\sum_{\alpha}\left|E_{\alpha}\right\rangle \left\langle E_{\alpha}\right|\otimes\hat{H}_{B;{\alpha}}$. Specifically, the state label takes the values $\alpha=S,-1,0,1$ for the isotropic exchange interaction and $\alpha=S,\widetilde{-1},\widetilde{0},\widetilde{1}$ with the dipolar interaction included. 
The nuclear Hamiltonians $\hat{H}_{B;{\alpha}}$ are also in general different between the two cases.
We consider an initial separable pure state $\left|\psi_S\right\rangle\otimes\left|\mathcal{J}\right\rangle$ of the full system at $t=0^+$, the electron spin part of which $\left|\psi_S\right\rangle$ is prepared by a first pulse at $t=0$. 
As discussed in Sect.~\ref{subsec:PCA_theory}, the pure dephasing dynamics is independent of the choice of the nuclear pure state $\left|\mathcal{J}\right\rangle$ for large baths, therefore evolution of this initial state well captures the dephasing dynamics of the completely mixed nuclear state simulated in gCCE. 

The evolution of the state is detailed in Supplementary Material Sect.~S4. In the case where isotropic exchange is the only interaction between the electron spins, as shown in Eq.~S7, the Hahn-echo pulse $\pi_{xx}=\exp[-i\pi(\hat{S}_{1x}+\hat{S}_{2x})]$ flips the states $\left|E_{-1}\right\rangle$ and $\left|E_{1}\right\rangle$ while essentially not affecting the $\left|E_{S}\right\rangle$ and $\left|E_{0}\right\rangle$ in terms of decoherence. This results in relatively simple expressions of the coherence functions $L^1_{\alpha\beta}(2\tau)$ in Eqs.~S10--S15, which is independent of the initial electron spin state $\left|\psi_S\right\rangle$. Each of the $L^1_{\alpha\beta}$ can be viewed as the overlap between two final nuclear bath states evolved through a pair of bifurcated paths in the state space under different $\hat{H}_{B;{\alpha}}$ Hamiltonian from the same initial $\left|\mathcal{J}\right\rangle$ state.

When the dipolar interaction is included between the electron spins, a complication arises since now the pulse $\pi_{xx}$ mixes the states $\left|E_{\widetilde{-1}}\right\rangle$, $\left|E_{\widetilde{0}}\right\rangle$ and $\left|E_{\widetilde{1}}\right\rangle$ states (Eqs.~S18--S21). Consequently, as shown in Eqs.~S32--S37, each of the $L^1_{\alpha\beta}$ now involves overlaps between nuclear states evolved through multiple different pairs of bifurcated paths from $\left|\mathcal{J}\right\rangle$ and is dependent on the initial electron-spin state, as the relative weight of the overlaps is controlled by the expansion coefficients $\beta_\alpha$ of $\left|\psi_S\right\rangle$ (see Eq.~S16). 
Furthermore, the projected nuclear Hamiltonians $\hat{H}_{B;{\alpha}}$ that determine these paths now depend on both the field strength $B$ and the dipolar interaction $D$ even in the leading order.
The stability of the gCCE method for calculating coherence functions with the full original Hamiltonian is also affected by the inclusion of the dipolar interaction and the resulting change in dynamics. 
Although in the FID scenario the gCCE still converges at order $1$, in the Hahn-echo experiment scenario it fails to converge for some of the coherence functions.
For a general orientation of $\vec{d}$ where $\phi\neq0$, although gCCE stills shows convergence behavior with respect to the order for $L^1_{\widetilde{-1},\widetilde{1}}$ as shown in Fig.~\ref{fig:dipolar_gcce_order_convergence}(a),
divergence with respect to the gCCE order and in terms of stronger fluctuations when the evolution time increases appears in $L^1_{\widetilde{-1},S}$, $L^1_{\widetilde{-1},\widetilde{0}}$, $L^1_{\widetilde{1},S}$ and $L^1_{\widetilde{1},\widetilde{0}}$ in a similar fashion. An example of $L^1_{\widetilde{-1},S}$ is shown in Fig.~\ref{fig:dipolar_gcce_order_convergence}(b).
$L^1_{S,\widetilde{0}}$ starts to show decay, although minor, on the decay time scale of $L^1_{\widetilde{-1},\widetilde{1}}$, and its gCCE calculation also diverges quickly with increasing time (Fig.~\ref{fig:dipolar_gcce_order_convergence}(c)). 
These results indicate the limitation of the gCCE method in simulating the decoherence of a system of spins coupled by anisotropic interactions under pulse sequences.

\begin{figure}[htp]
$\begin{array}{c}
\includegraphics[width=0.95\linewidth]{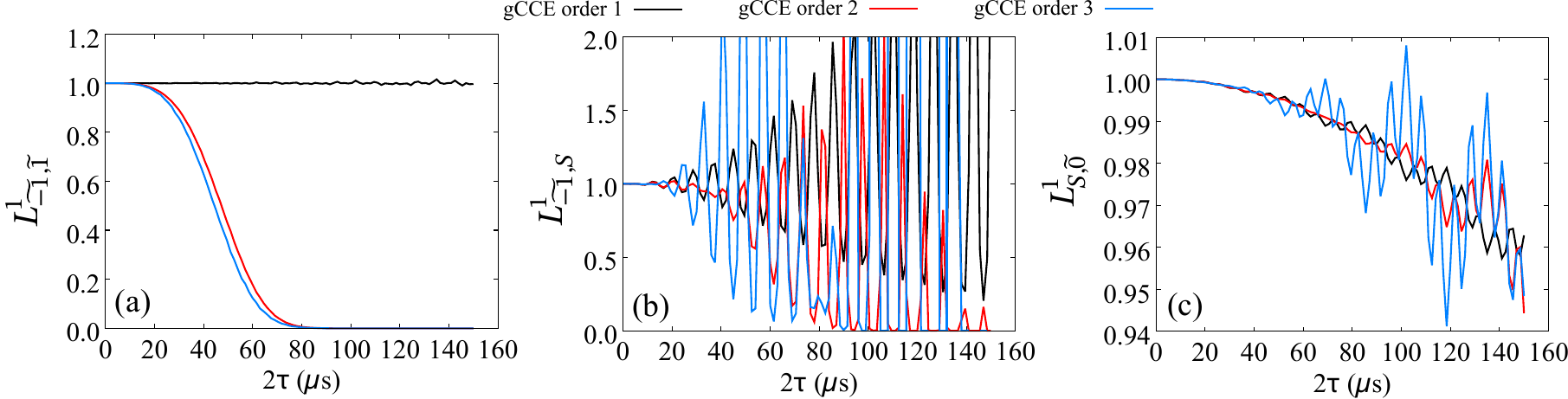}
\end{array}$
\caption{{\color{black}
gCCE calculated coherence functions (a) $L^1_{\widetilde{-1},\widetilde{1}}$, (b) $L^1_{\widetilde{-1},S}$, (c) $L^1_{S,\widetilde{0}}$ at different orders for a single random spatial configuration of protons.
Parameters of the model used for results in this figure are $\phi=45\degree$, $d=10\ang$ ($D=52.04\MHz$), $J=80\MHz$, $B=0.5\T$, $R_S=20\ang$, $R_B=2\ang$, $n_B=0.01/\ang^{3}$.}} \label{fig:dipolar_gcce_order_convergence}
\end{figure}

}

\section{Conclusions}\label{sec:Conclusion}

In this work, we consider a model of two electron spin-$\frac{1}{2}$'s  interacting via an isotropic exchange interaction and are
embedded in a random proton spin bath. 
The decoherence of the two-spin system for both the scenarios of FID and Hahn-echo is studied, and
we focus on model parameters in the range mimicking typical magnetic molecules and on the pure dephasing regime.
The coherence times $T_2^*$, for FID and $T_2$, for Hahn-echo, of the off-diagonal RDM elements in the energy basis are extracted from the gCCE simulation of the time evolution. 
The dependence of $T_2^*$ and $T_2$ on each of the model parameters of external magnetic field orientation and strength, magnitude of the exchange interaction, distance between the two electron spins, distance between the electrons and the closest proton spins, proton density and minimum spacing among protons is investigated with most other parameters fixed. 
We find optimal parameter ranges for longer coherence times{\color{black}, provide physical understanding of these ranges with the help of the PCA method} and discuss implications for experiments.
{\color{black}
The changes in the dephasing dynamics and in the stability of the gCCE method resulting from including a non-negligible magnetic dipolar interaction between the two electron spins are also discussed. 
}
Our work provides useful insights on how various factors can be optimized to enhance the coherence time of two coupled electron spins in magnetic molecules, which is crucial for realizing high-fidelity two-qubit entangling gates as well as single qubits encoded in two-spin systems.

\section{Supplementary Material}
See Supplementary Material for (1) the description of the generalized cluster correlation expansion method, (2) the computational details, (3) tests of convergence with respect to the gCCE order, and (4) a derivation of the evolution of the full electron-nuclei spin pure state in the Hahn-echo experiment and the electron spin coherence functions expressed in terms of the pure state, for both isotropic exchange and dipolar interactions between the two electron spins.

\begin{acknowledgments}
The authors are grateful for useful conversations with Shuanglong Liu, Haechan Park, and Steve Hill. This work is supported by the U.S. Department of Energy, Office of Science, Basic Energy Sciences under Award No.{} DE-SC0022089. The calculations were performed using the utilities of the National Energy Research Scientific Computing Center and the University of Florida Research Computing.
\end{acknowledgments}

%\section{AUTHOR DECLARATIONS}
\section*{Author Declarations}

\textbf{Conflicts of Interest: } The authors have no conflicts of interest to disclose.

\textbf{Author Contributions: } Xiao Chen: Conceptualization (equal); Data Curation (lead); Formal Analysis (lead); Methodology (lead); Visualization (lead); Writing/Original Draft Preparation (lead); Writing/Review and Editing (equal). Silas Hoffman: Conceptualization (equal); Formal Analysis (supporting); 
%Project Administration (equal); 
Visualization (supporting); Writing/Original Draft Preparation (supporting); Writing/Review and Editing (equal). James N. Fry: Writing/Review and Editing (equal). Hai-Ping Cheng: Conceptualization (equal); Funding Acquisition (lead); Project Administration (lead); Supervision (lead); Writing/Review and Editing (equal).

\section*{Data Availability Statement}

The data that support the findings of this study are available from the corresponding author upon reasonable request.

\bibliography{main.bib}

\end{document}